\documentclass[a4paper,12pt]{article}
\usepackage{soul}
\usepackage[usenames,dvipsnames]{color}
\usepackage{jheppub}
\usepackage{esvect}
\usepackage{amsmath, amssymb, slashed, epsf, color, graphicx, latexsym, tensor, physics, xcolor, array, tabularx, makecell}
\usepackage{epsfig}
\usepackage{comment}
\usepackage{graphics}
\usepackage{mathtools}
\usepackage{appendix}
\usepackage{inputenc}
\usepackage{indentfirst}

\title{\boldmath Quantum evolution of de Sitter black holes near extremality}
\author[1]{Arindam Bhattacharjee,}
\author[2]{Muktajyoti Saha}

\affiliation[1]{Faculty of Physics, University of Warsaw, \\
ul. Pasteura 5, 02-093 Warsaw, Poland}
\affiliation[2]{International Centre for Theoretical Sciences - TIFR, Bengaluru - 560089, India}

\emailAdd{arindamb.hep@gmail.com}
\emailAdd{muktajyoti.hepth@gmail.com}

\abstract{We study the evolution of charged, asymptotically de Sitter black holes close to the cold extremal branch of the phase space. We consider black hole sizes that are parametrically smaller than both their inverse temperature and the cosmological horizon. Unlike flat space, charged de Sitter black holes do not evolve towards extremality, but rather towards a thermal equilibrium with the cosmological horizon. In the low-temperature regime, the near-horizon physics can be effectively captured by a one-dimensional Schwarzian theory. This is coupled to the far-horizon de Sitter quantum field theory. Incorporating the thermal nature of the cosmological horizon, we compute the quantum energy transfer through uncharged massless scalar particles. The results significantly differ from Hawking's thermal predictions. Black holes that are hotter than the cosmological horizon emit energy at a rate lower than their asymptotically flat counterparts. Whereas much colder ones absorb energy at a nearly constant rate.
}

\begin{document} 
\maketitle

\section{Introduction}

In classical gravitational theories, black holes are certain solutions characterized by mass, charges, and angular momenta. They have a singularity hidden behind a horizon, from which nothing can escape. This is no longer true if semiclassical effects are taken into account. Hawking showed that black holes radiate thermal energy at a temperature given by their surface gravity. They also carry entropy, given by the area of the horizon. In asymptotically flat spacetimes, Hawking radiation takes the charged and rotating black holes towards an `extremal limit' where the temperature becomes zero. At this limit, the charge/angular momentum parameter of a black hole takes its maximum allowed value for a given mass. Non-extremal black holes generically have two horizons, and at extremality, these two horizons coincide. Asymptotically flat extremal black holes are characterized by their near-horizon geometry, which contains an infinitely long AdS$_2$ throat that governs their physics \cite{Sen:2007qy, Sen:2008vm}. They also have infinite-dimensional asymptotic symmetries emerging from the AdS$_2$ factor.

Near-extremal black holes are solutions that are slightly deviated away from extremality. They have small nonzero temperatures, and their near-horizon geometry has finite but very large throats. In this regime, the semiclassical black hole energy is even lower than the typical energy of a Hawking quantum at that temperature, making the semiclassical picture of evaporation questionable \cite{Preskill:1991tb}. Thus, the semiclassical description of these black holes fails, and quantum gravitational effects become more important. Unlike typical black holes, these quantum effects are better understood owing to recent progress in the literature. These quantum effects can be effectively described in terms of one-dimensional modes localized in the near-horizon boundary \cite{Nayak:2018qej, Moitra:2019bub, Iliesiu:2020qvm, Heydeman:2020hhw, Sachdev:2019bjn, Larsen:2018iou, Castro:2018ffi}. From the four-dimensional perspective, these dynamical modes are a consequence of the soft breaking of certain asymptotic symmetries of the near-horizon AdS$_2$ of the corresponding extremal black hole \cite{Iliesiu:2022onk, Banerjee:2023quv, Banerjee:2023gll}. Among these symmetries, there is a universal Schwarzian sector associated with the breaking of large diffeomorphisms of AdS$_2$. Depending on the isometries and matter gauge fields, usually, there can be additional gauge modes as well. These broken symmetries significantly modify the thermodynamics, such as their entropy gets logarithmic temperature contributions \cite{Banerjee:2023quv, Banerjee:2023gll, Kapec:2023ruw, Rakic:2023vhv, Maulik:2024dwq, Modak:2025gvp}. Consequently, the evaporation dynamics of such black holes also deviate from Hawking's semiclassical prediction of a thermal radiation spectrum \cite{Brown:2024ajk, Bai:2023hpd, Maulik:2025hax, Lin:2025wof}. The effects of the Schwarzian modes in various facets of near-extremal black hole physics have been discussed in \cite{Kapec:2024zdj, Kolanowski:2024zrq, Emparan:2025qqf, Emparan:2025sao, Biggs:2025nzs, Betzios:2025sct, Liu:2024gxr, Heydeman:2024ezi, Li:2025vcm, Castro:2025itb}.

So far, we have talked about asymptotically flat black holes. However, at the largest scales, the current epoch of our universe can be described as having a small positive cosmological constant. de Sitter spacetime is a maximally symmetric solution in Einstein gravity with positive cosmological constant. Thus, understanding the quantum gravitational effects in the evolution of asymptotically de Sitter black holes is an interesting problem. The charged or rotating de Sitter black hole phase space is much richer than that of their asymptotically flat counterparts. In this paper, we will talk about the Reissner-Nordstr\"om-dS solution for simplicity. This solution has two black hole horizons and a larger cosmological horizon. Due to this, there are three extremal limits: Cold, Nariai, and Ultracold. Thus, the study of near-extremal black holes in de Sitter is also more involved. The Schwarzian-type modes originating from the asymptotic symmetries are present near the cold extremal limit, where the near-horizon geometry has an AdS$_2$ factor. These modes give rise to similar logarithmic temperature contributions to entropy \cite{Maulik:2025phe, Blacker:2025zca, Mariani:2025hee, Arnaudo:2025btb}. Infinite-dimensional near-horizon symmetries also emerge close to the other two extremal limits, but these are less understood\footnote{See \cite{Chen:2025lnk} for recent discussions on near-Nariai black holes.} \cite{Blacker:2025zca}.

Furthermore, the cosmological horizon also has thermal properties, and it radiates at a temperature set by the characteristic de Sitter length scale. Thus, black holes in de Sitter evolve towards a thermal equilibrium with the cosmological horizon \cite{Montero:2019ekk}. In fact, extremal black holes in de Sitter should absorb energy. This is unlike flat space, where black holes evolve towards zero temperature or extremality. In this paper, we take into account these thermal effects of dS cosmological horizon in understanding the evolution of charged near-extremal black holes in de Sitter. We consider a small near-extremal black hole close to the cold extremal limit. The \textit{near-extremal} condition $r_0\ll \beta$ ensures that the near-horizon geometry is very large, and Schwarzian-type modes are present. Here $\beta$ and $r_0$ are the inverse temperature and the extremal horizon size. The \textit{smallness} $r_0\ll L$ ensures that the far-horizon region is also large, where the spacetime can be approximated as pure de Sitter. Here, $L$ is the de Sitter length scale. We adopt the ideas of \cite{Brown:2024ajk} in coupling the near and far-horizon systems along the near-horizon boundary. From the perspective of an observer in the far-horizon region, the near-extremal black hole can then be modelled as a Schwarzian system in the deep-bulk, interacting with the ambient de Sitter space. The new ingredient of our computation is to appropriately incorporate the thermal nature of the cosmological horizon. We compute the energy transfer rates using time-dependent perturbation theory. These rates are very different from Hawking's semiclassical prediction of a thermal spectrum. We find that when the black holes are hotter than the cosmological horizon, the net emission rate is lower than that in asymptotically flat spacetime. This is because the black hole is radiating into a thermal bath. We further find that black holes that are much colder than the cosmological horizon absorb energy at an almost constant rate.   

The content of the paper is as follows: In section \ref{sec:dS-basics}, we review some important aspects of empty de Sitter spacetime, including the classical geometry and the basics of quantum fields in de Sitter background. In section \ref{sec:dSBH}, we discuss the phase space of charged de Sitter black holes, with a focus on near-extremal black holes. We then consider massless scalar field in near-extremal background and write down the semiclassical greybody factors for arbitrary angular momenta, slightly extending known results. Section \ref{sec:BH-evol} contains the main calculations of the paper. Here, we compute the quantum evolution rates using an effective Schwarzian description giving us the energy dissipation rate \eqref{quantum-rate}. We compare our results with the semiclassical predictions and find significant deviations for black holes with low enough energies. We also contrast the evolutions of black holes that are respectively hotter and colder than the lukewarm line. We summarize our results in section \ref{sec:disc}. The appendices \ref{app:static} and \ref{app:id} contain some computational details.

\section{de Sitter basics}\label{sec:dS-basics}
de Sitter spacetime is the maximally symmetric solution of Einstein equations with a positive cosmological constant. Our current understanding of the large scale structure of spacetime postulates a very small but positive cosmological constant present in our universe. Hence understanding gravitational dynamics in de Sitter background is crucial. Below we recall basics of de Sitter geometry.
\subsection{Coordinate charts}
Empty de Sitter space of four dimensions ($dS_4$) is a hypersurface embedded in $(4+1)$ dimensional Minkowski space $\mathbb{M}_{4+1}$ with the following relation:
\begin{align}
    -X_0^2+\sum_i X_i^2 = L^2,
\end{align}
where $\{X_0,X_i\}$ with $i\in \{ 1,...,4\}$ are the temporal and spatial coordinates. $L$ denotes the radius of the de Sitter space, which is related to the cosmological constant via,
\begin{align}
    \Lambda = \frac{3}{L^2}.
\end{align}
The spatial sections are hyperspheres $\mathbb{S}^3$, which becomes clear as we look at the global coordinates that cover the whole de Sitter space. The de Sitter metric in this coordinate takes the form,
\begin{align}
    ds^2 = -d\tau^2 + L^2 \cosh^2\left(\frac{\tau}{L}\right) d\Omega_3^2,
\end{align}
where $d\Omega_3^2$ is the standard metric on a unit 3-sphere $\mathbb{S}^3$. The isometry group of this metric is $SO(4,1)$, which is inherited from the embedding flat space. This also clarifies the de Sitter radius $L$ as being the radius of the extremal $\mathbb{S}^3$ at $\tau =0$. The radius of the spatial sphere increases both towards the future and the past of $\tau =0$ and becomes infinite as $\tau \xrightarrow{} \pm \infty$. Our universe is currently expanding, which can be modeled by the $\tau >0$ part of the de Sitter spacetime.

However, we must also understand de Sitter spacetime from the point of view of the observers who live there. For a static observer, living in say the north pole of the spatial $\mathbb{S}^3$, the whole of the spatial section is not available to probe. So, we can define the `causal diamond' for such an observer, which is the part of spacetime they can send and receive information from. This region is called the static patch of de Sitter and is covered by the static coordinates. The metric is given by,
\begin{align}
    ds^2 = -\left(1-\frac{r^2}{L^2}\right) dt^2 + \frac{dr^2}{\left(1-\frac{r^2}{L^2}\right)} + r^2 d\Omega_2^2,
\end{align}
where $t \in \{-\infty,\infty\}$ and $r\in \{0,L\}$ is the region inside the static patch and $\Omega_i$ are the coordinates of 2-sphere. The Penrose diagram of de Sitter is shown below in figure \ref{fig:dSpen},
\begin{figure}[htbp]
    \centering
    \includegraphics[width=0.7\textwidth]{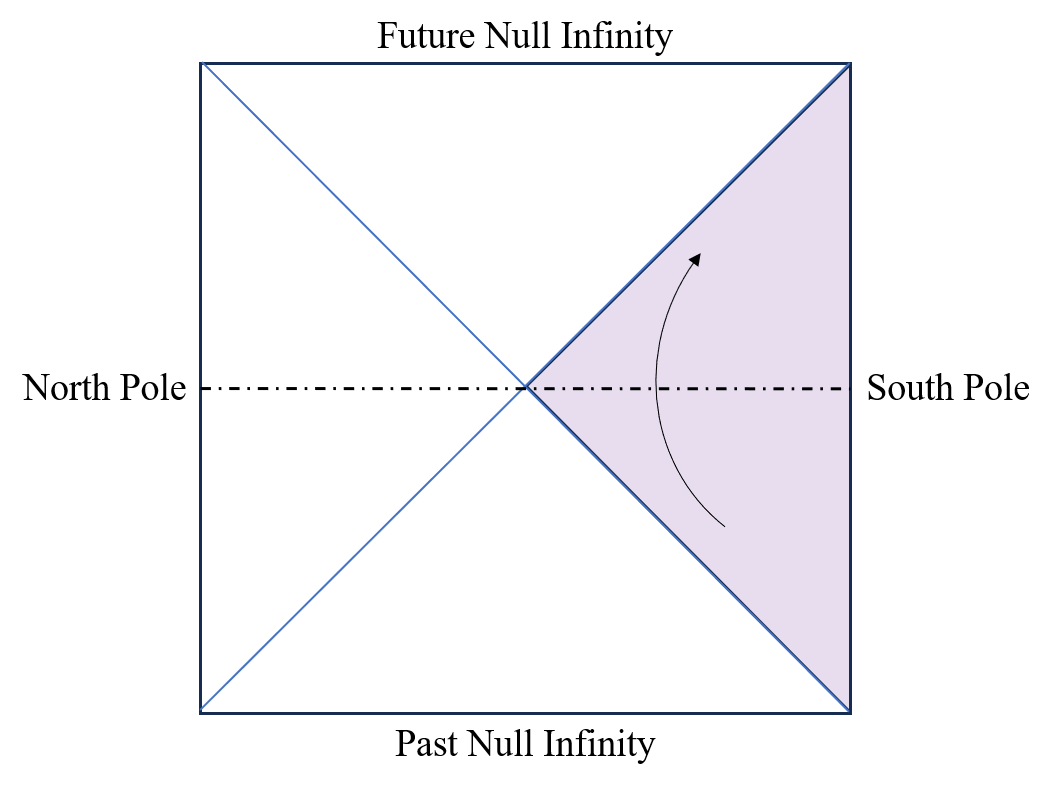}
    \caption{de Sitter Penrose diagram. The dashed line shows the spatial section of global de Sitter which is $\mathbb{S}^3$. The highlighted area is the static patch for an observer at south pole. The static time arrow is also shown.}
    \label{fig:dSpen}
\end{figure}

Firstly, in these coordinates, the spatial slices are non-compact planes rather than compact $\mathbb{S}^3$, which changes the quantisation properties of fields as we will discuss below. Secondly, in these coordinates, we have a timelike Killing vector $\partial_t$, so we can sensibly define a conserved Hamiltonian and study its evolution. We also observe that at $r=L$, there exists a horizon. The norm of $\partial_t$ becomes zero at $r=L$ so it is also a Killing horizon. This is the cosmological horizon, which is absent in flat space, and the presence of this horizon makes the study of de Sitter black holes much richer than their flat space analogues.

\subsection{QFT and different vacua in dS}\label{subsec:dS-qft}

Let us briefly comment on quantum fields in de Sitter background. A general property of quantum fields in curved spacetime is that there are no unique definitions of vacua. It depends on the choice of the observer. Since we are interested in the static observer, let us first take their example. For these observers, the static time translation is a symmetry and hence the killing vector $\partial_t$ can be used to define positive and negative `frequency' modes $\{v^{\pm,\omega}_S\}$. The explicit expression for these modes is given in section \ref{sec:dSBH}. In particular, equations (\ref{phi-mode-exp}) along with (\ref{FHR-sol}) give an explicit construction of such modes. Thus, we can proceed in the usual way and define a `Static vacuum' $\ket{0}_S$ by defining creation and annihilation operators and demanding $\ket{0}_S$ is annihilated by all annihilation operators \cite{Higuchi:1986ww}.

However, the vacuum defined as above will not be invariant under the isometries of the de Sitter group $SO(4,1)$. Thus, the choice of Static vacuum will explicitly break the classical symmetries in the QFT. Hence, another natural choice for a vacuum would be the one that preserves the isometries of the de Sitter background. Such a vacuum, termed `Bunch-Davies vacuum', does exist \cite{Bunch:1978yq} and is defined through a different sets of modes $\{v_{BD}^{\pm,k} \}$ \footnote{In fact, there is a one-parameter family of dS-invariant vacua, named the $\alpha$-vacua \cite{Allen:1985ux}. For a nice review, see \cite{Chopping:2024oiu}.}.

These vacua are not equivalent. This can be seen from the fact that the positive frequency modes of one set cannot be expressed as a linear combination of the positive frequency modes of the other set. In fact, they have the following relation by analytic continuation \cite{Higuchi:2018tuk},
\begin{align}\label{global-static-mode-reln}
    v_S^{+,\omega}(x) = \frac{1}{\sqrt{1-e^{-2\pi \omega L}}}(v_{BD}^{+,\omega} + e^{-\pi \omega L}v_{BD}^{-,\omega}).
\end{align}
This ultimately implies that the global vacuum manifests itself as a `thermal state' to a static observer. This is the quantum version of the classical statement that the cosmological horizon of pure de Sitter space has a temperature given by $T_c=\frac{1}{2\pi L}$. For us, the state $\ket{0}_{BD}$ is precisely the thermal bath which will interact with our black hole.

Apart from the general problem of defining a vacuum state, massless scalar fields in de Sitter suffer from infrared divergence issues. Since we will consider radiation from the de Sitter black holes through massless fields, we need to address these concerns. We will follow the regularization procedure of adding a small mass `$m$' to the field as an IR regulator and finally take $m\xrightarrow{}0$ limit to get the expressions for the massless fields (see \cite{Alicki:2023rfv} for example). It has been argued that these kinds of infrared divergences for generic massless fields are absent when appropriate observables are considered or under physical gauge choices \cite{Page:2012fn, Bernar:2014lna}.

\section{de Sitter black holes}\label{sec:dSBH}
We now discuss black hole solutions in gravitational theories with a positive cosmological constant. Typical examples are Schwarzschild-dS, Reissner-Nordstr\"{o}m-dS, and Kerr-dS. In this paper, we will focus on Reissner-Nordstr\"{o}m-de Sitter black holes, which are static black hole solutions of Einstein-Maxwell theory with positive cosmological constant. Below, we discuss the phase space of these black holes.

\subsection{Charged black hole phase space}\label{subsec:phase}
\noindent In static coordinates, the black hole metric takes the form,
\begin{align}\label{met-RN}
    ds^2 = - f(r)dt^2 + \frac{dr^2}{f(r)} + r^2 d\Omega_2^2,
\end{align}
with $f(r) = 1-\frac{r^2}{L^2}-\frac{2M}{r}+\frac{Z^2}{r^2}$, where $M$ is the mass parameter of the black hole and $Z$ is the charge associated with the background gauge field,
\begin{align}
    Z^2 = Q^2 + P^2 \quad\text{where} \quad A = \frac{Q}{r} dt - P \cos \theta d\phi.
\end{align}
We will assume electrically charged black holes with $P = 0 \implies Z=Q$. We have set the Newton constant to 1.

\begin{figure}[htbp]
    \centering
    \includegraphics[width=0.7\textwidth]{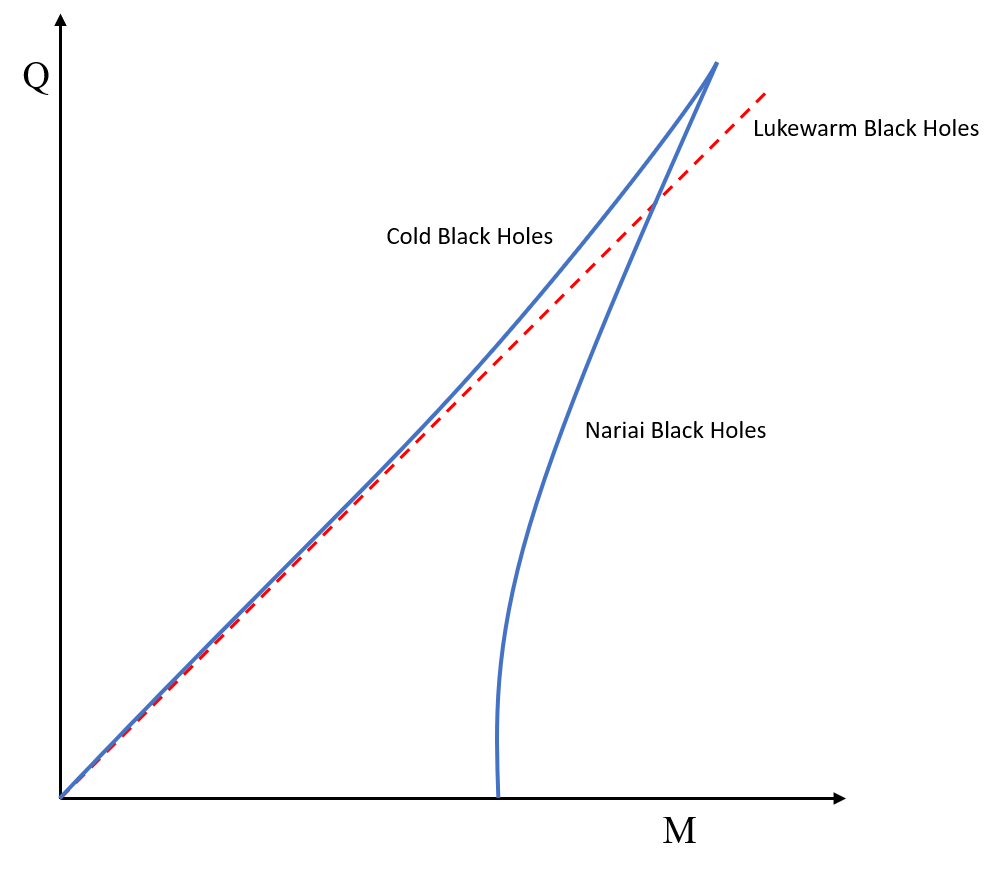}
    \caption{A Schematic diagram of electrically charged dS black hole phase space}
    \label{fig:shark}
\end{figure}

This metric approaches the de Sitter metric at large $r$. Although the only physical singularity, as can be seen from curvature invariants, is at $r=0$, there exist coordinate singularities. These coordinate singularities give rise to horizons,  located at $f(r) = 0$. In general, there will be three positive roots, which we call the inner horizon ($r_-$), outer horizon ($r_+$), and cosmological horizon ($r_C$), in ascending order, i.e., $r_- < r_+ < r_C$. In terms of the roots, $f(r)$ takes the following form,
\begin{align}\label{fr-form}
    f(r) = -\frac{1}{L^2 r^2} (r-r_-)(r-r_+)(r-r_C)(r + r_- + r_+ + r_C).
\end{align}
The relations between the roots and the parameters are given by,
\begin{align}
    & M = \frac{1}{2L^2} (r_{+} + r_{-})(L^2 - r_{+}^2 - r_{-}^2), \nonumber\\
    & Q^2 = \frac{r_{+}r_{-}}{L^2}(L^2 - r_{+}^2 - r_{-}^2 - r_{+}r_{-}), \nonumber\\
    & L^2 = r_C^2 + r_{+}^2 + r_{-}^2 + r_{+}r_{-} + r_C(r_{+} + r_{-}).
\end{align}
But these roots are not real for generic values of the parameters $\{M,Q\}$. This is analogous to flat space charged black holes where an `extremal' limit exists, beyond which the horizons become complex, implying a naked singularity.
Because Reissner-Nordstr\"{o}m black holes have three horizons, there are three distinct extremal limits possible for these black holes:
\begin{itemize}
    \item When $r_-$ and $r_+$ approach each other, i.e. $r_-=r_+ = r_0< r_C$. These are called `Cold Black Holes'.
    \item When $r_+$ and $r_C$ approach each other i.e. $r_-<r_+=r_C$. They are called Nariai Black holes.
    \item Finally when all three horizons approach each other $r_-=r_+=r_C$. They are called Ultracold black holes.
\end{itemize}
This is evidenced by the fact that, unlike flat space where extremality implies a simple $M=Q$ relation between the two parameters, in de Sitter space the corresponding critical curve is $(-27 M^4-8 Q^4+36 M^2 Q^2)\frac{1}{L^2}+M^2-16 Q^6-Q^2 = 0$. This gives the famous shark-fin diagram \ref{fig:shark} in the phase space.

The usual laws of Black hole thermodynamics applies to each of these horizons. At each horizon $r_h$ the temperature and entropy are defined as follows:
\begin{align}\label{temp-bh-defn}
	T_h = \frac{1}{2\pi}\abs{f'(r_h)} \qquad S = \pi r_h^2.
\end{align}

\subsubsection*{Cold Black holes}
Clearly, among the three extremal branches, only the cold black holes have flat space analogues. So, let us first focus on the extremal black holes in this case. Setting $r_+ = r_- = r_0$ in \eqref{fr-form} and comparing with the form of $f(r)$ given above, we see  
\begin{align}
    & M_0 = \frac{r_0}{L^2}  (L^2 - 2 r_{0}^2), \\
    & Q_0^2 = \frac{r_{0}^2}{L^2}(L^2 - 3 r_{0}^2 ), \\
    & L^2 = r_C^2 + 3 r_{0}^2 + 2 r_C r_{0}.
\end{align}

\subsubsection*{Lukewarm Black Holes}
Apart from the extremal limits in the phase space, one more interesting limit that exists is the Lukewarm black holes. They are defined by the black holes for which the temperature of the outer horizon ($r_+$), and the cosmological horizon ($r_C$) are the same. They lie in the $M=Q$ line in the phase space. In the absence of any charged particle emission, these lukewarm black holes are in thermodynamic equilibrium, and they are the endpoint of all black hole evolution \cite{Montero:2019ekk}. For small black holes, we can show\footnote{We are considering evolution through uncharged particles only, hence we move along a fixed charge line. In presence of charged particles, the only equilibrium point is the empty de Sitter spacetime.},
\begin{align}
	M_{lw} = M_0 + \frac{r_0^3}{2L^2} \qquad Q_{lw} = Q_0
\end{align}
Any black hole with temperature above $T_{lw}$ would flow towards the lukewarm line by net emission of energy. The external observer will see this as the Hawking radiation. Whereas, any black hole that has temperature below $T_{lw}$ would absorb net energy from the surrounding de Sitter spacetime. This later phenomenon is completely absent in flat space black holes, and it makes the study of de Sitter charged black holes extremely exciting.

\subsection{Near-extremal Black holes}
We want to study the evolution of near-extremal, small electrically charged black holes in de Sitter. The phase space of these near-extremal black holes is analysed in \cite{Castro:2022cuo}.
Generically, an extremal black hole is characterised by universality in the near-horizon geometry. For cold black holes in de Sitter, the near-horizon geometry takes the form $AdS_2 \times \mathbb{S}^2$, where the $AdS_2$ radius $l_{AdS}$ is given by
\begin{align}\label{ads2-radius}
l^2_{AdS} = \frac{r_0^2}{(1-6\frac{r_0^2}{L^2})} \approx r_0^2 (1+ 6\frac{r_0^2}{L^2}) \quad \text{for small BHs}
\end{align}
and $\mathbb{S}^2$ has radius $r_0$. The near-horizon field strength also takes a constant value proportional to the volume 2-form of the AdS$_2$ throat.

To analyse black hole solutions close to extremality, we consider a very small temperature $T$ such that $r_0 T \ll 1$. In the phase space, we move away from extremality by starting from an extremal black hole $\{M_0,Q_0\}$ and then slightly changing the mass to
\begin{align}
M = M_0 + T^2 \frac{2\pi r_0^3 }{(1-6\frac{r_0^2}{L^2})}+\cdots
\end{align}  
while keeping $Q = Q_0$. From the expansion above, we can define an energy scale 
\begin{align}
E_0 = \frac{(1-6\frac{r_0^2}{L^2})}{2\pi r_0^3 }.
\end{align}
This represents the typical energy scale above extremality, where the semiclassical picture of black hole evaporation breaks down \cite{Preskill:1991tb}. In the extremal geometry, the horizons $r_+$ and $r_-$ merge, but for near-extremal black holes, they are slightly separated, 
\begin{align}
r_{\pm} = r_0 \pm 2\pi l_{AdS}^2 T + \mathcal{O}(T^2).
\end{align}
Even for near-extremal black holes, the geometry can be written as $AdS_2 \times \mathbb{S}^2$ but with finite temperature corrections to this product geometry. This is best seen in a two-dimensional picture. Using the symmetries of the near-horizon geometry, the Einstein-Maxwell theory of the full spacetime can be dimensionally reduced along the compact $\mathbb{S}^2$ space. This is a standard procedure, and if we concentrate only on the massless sector, it gives rise to a dilaton gravity theory coupled to a gauge field in $AdS_2$.

After non-trivial manipulations exploiting the form of the near-extremal corrections to the near-horizon throat, the effective theory ultimately boils down to a Schwarzian theory (coupled to additional one-dimensional modes) living at the boundary of the $AdS_2$. This effective theory governs the dynamics of the low-temperature near-extremal black holes \cite{Nayak:2018qej, Iliesiu:2020qvm}. With appropriate boundary conditions on the boundary of $AdS_2$, this dual theory provides us with quantum-corrected density of states for the black hole system. For a fixed charge $Q$ and energy above extremality $E= M-M_0$, the density of states is given by 
\begin{align}\label{sch-dos}
	&\rho_{\text{Sch}}(E) = \frac{e^{S_0}}{2\pi^2 E_0}\sinh(2\pi\sqrt{2E E_0^{-1}})
\end{align}
This corrects the naive semiclassical counting of black hole states for near-extremal black holes. Since we are interested in black hole evolution, we will also need to understand transitions in the Schwarzian picture. The transitions between Schwarzian states will be governed by certain operators sourced by the far-horizon quantum field. For a generic operator of conformal dimension $\Delta$, the matrix elements between two energy states are given by \cite{Mertens:2018fds},
\begin{align}\label{sch-corr}
	& \abs{\bra{E_f}{O}_{\Delta}\ket{E_i}}^2 = \frac{2e^{-S_0}}{(2E_0^{-1})^{2\Delta}\Gamma(2\Delta)}\prod_{\pm}\Gamma \left(\Delta\pm i\sqrt{2E_i E_0^{-1}})\pm i\sqrt{2E_f E_0^{-1}}\right)
\end{align}
We will extensively use this picture to understand the evolution of these black holes.

\subsection{Massless scalar field in dS BH background}

Now, we consider a massless scalar field in the Reissner-N\"ordstrom-dS (RNdS) background \eqref{met-RN} satisfying the Klein-Gordon equation of motion $\nabla^2\Phi = 0$. In general, it is difficult to solve the equation in the full background. However, as we are considering a small near-extremal black hole, the spacetime can be divided into two large asymptotic regions: a near-horizon region, where the geometry is dominated by the black hole potential, and a far-horizon region, where the influence of the cosmological constant becomes significant. The scalar field solutions in these two regions can be obtained independently and subsequently matched in an intermediate overlap region \cite{Harmark:2007jy}. We consider the ansatz,
\begin{align}\label{phi-mode-exp}
    & \Phi(t,r,\Omega) = \text{e}^{-i \omega t} Y_{lm}  (\Omega) v_{\omega l}(r).
\end{align}
We also consider the Tortoise coordinate defined through the relation $\frac{dx}{dr} = f(r)^{-1}$. In these coordinates, the equation takes the form of a Schr\"odinger equation:
\begin{align}\label{schrodinger-bh}
    \left(\frac{d^2}{dx^2} + \omega^2 - V(r) \right)(r v_{\omega l}) = 0, \quad V(r) = \frac{f(r)}{r^2}(l(l+1) + r f'(r)),
\end{align}

\subsubsection*{Solution in near-horizon region (NHR)}
We start with the near-horizon region of the background. In particular, we consider the region $r_{+}<r \ll L$ while the horizon radii satisfy $r_{\pm} \ll L$. Under these limits, the $g_{tt}$ component becomes exactly like the near-horizon limit of RN solution in flat space,
\begin{align}
    f(r) \approx \frac{(r - r_{+})(r - r_{-})}{r_{+}^2}.
\end{align}
Hence, the effects of de Sitter space become negligible. The scalar field solution is given by,
\begin{align}
    v_{\omega l}(r)|_{\text{NHR}} = c P_{l}^{-i\beta\omega/2\pi}\left(-1-\frac{\beta (r - r_{+})}{2\pi r_{+}^2}\right) + c' Q_{l}^{-i\beta\omega/2\pi}\left(-1-\frac{\beta (r - r_{+})}{2\pi r_{+}^2} \right).
\end{align}
Here, $\beta$ is the inverse temperature associated with the black hole outer horizon. $P_l^m, Q_l^m$ are the associated Legendre polynomials of first and second kinds respectively. Now, we consider the asymptotic behaviors of the solutions. Near the horizon we have:
\begin{align}
    & P_{l}^{-i\beta\omega/2\pi}\left(-1-\frac{\beta (r - r_{+})}{2\pi r_{+}^2}\right) \xrightarrow[r\to r_{+}]{} (r - r_{+})^{-i\beta\omega/4\pi}, \\
    & Q_{l}^{-i\beta\omega/2\pi}\left(-1-\frac{\beta (r - r_{+})}{2\pi r_{+}^2} \right) \xrightarrow[r\to r_{+}]{} (r - r_{+})^{\pm i\beta\omega/4\pi}. 
\end{align}
Thus, the $P_l^m$-mode is purely ingoing at the horizon, while the $Q_l^m$-mode has both ingoing and outgoing waves. Hence, to put a purely ingoing boundary condition at the horizon, we set $c' = 0$.

Now, we consider the $r\to \infty$ limit of the solution. This brings us to the overlap of near-horizon and far-horizon regions. The falloff is given by,
\begin{align}\label{BH-sol-match}
    v_{\omega l}(r)|_{\text{NHR}} \xrightarrow[r\to \infty]{} c (b_1 r^l + b_2 r^{-l-1}).
\end{align}
From the asymptotic expansion of the associated Legendre polynomial, $b_1,b_2$ are fixed coefficients, given by:
\begin{align}
    & b_1 =  (-1)^l \text{e}^{\beta\omega/4}\pi^{-\frac{1}{2}-l}r_{+}^{-2l}\beta^l \frac{\Gamma\left(\frac{1}{2} + l\right)}{\Gamma\left(l+1+\frac{i\beta\omega}{2\pi}\right)}, \\
    & b_2 =  (-1)^l \text{e}^{\beta\omega/4}\pi^{\frac{l}{2}+l}r_{+}^{2+2l}\beta^{-l-1} \frac{\Gamma\left(-\frac{1}{2} - l\right)}{\Gamma\left(-l+\frac{i\beta\omega}{2\pi}\right)}.
\end{align}

\subsubsection*{Solution in far-horizon region (FHR)}
Next, we consider the solution in far-horizon region. Here, $r_{\pm}\ll r < L$. In this region, the metric appears to be pure de Sitter spacetime in static patch. The $g_{tt}$ component is given as,
\begin{align}
    f(r) \approx 1 - \frac{r^2}{L^2}.
\end{align}
The solution in this region is given by,
\begin{align}\label{FHR-sol}
    & v_{\omega l}(r)|_{\text{FHR}} = a_1 L^{-l-1}r^{l} \left(1 - \frac{r^2}{L^2}\right)^{-iL\omega/2} F\left(\frac{l-iL\omega}{2},\frac{3+l-iL\omega}{2};\frac{3}{2}+l;\frac{r^2}{L^2}\right) \nonumber\\
    & + a_2 L^l r^{-l-1} \left(1 - \frac{r^2}{L^2}\right)^{-iL\omega/2} F\left(\frac{-1-l-iL\omega}{2},\frac{2-l-iL\omega}{2};\frac{1}{2}-l;\frac{r^2}{L^2}\right).
\end{align}
Here, $F$ denotes the Hypergeometric function ${}_2F_1 (a,b;c;z)$. We consider the asymptotic behavior $r\to 0$ of the solution, which takes us to the overlapping region,
\begin{align}\label{FHR-CH-lim}
    v_{\omega l}(r)|_{\text{FHR}}\xrightarrow[r\to 0]{} a_1  L^{-l-1} r^l + a_2 L^l r^{-l-1}.
\end{align}
Matching this solution with the asymptotic limit \eqref{BH-sol-match} of the NHR solution, we obtain:
\begin{align}\label{match-cond}
    a_1 = c L^{l+1} b_1, \quad a_2 = c L^{-l} b_2.
\end{align}
Note that in pure dS spacetime, the mode proportional to $a_2$ diverges as $r\to 0$, rendering it unphysical within the static patch. Consequently, regularity at the origin requires setting 
$a_2 = 0$, leaving a single physically admissible solution that remains finite throughout the region $0<r<L$. This regular mode can be analytically continued beyond the cosmological horizon into the future region $L<r<\infty$, where it extends to form a complete solution over the Poincaré patch of de Sitter spacetime. The Poincaré patch description plays a crucial role in defining the de Sitter-invariant Bunch-Davies vacuum, which serves as the natural ground state for quantum field theory in de Sitter. In contrast, due to the presence of a horizon in the RNdS background, there is no such regularity issue. The interpretation of the $a_2$ mode corresponds to tunneling out of the near-horizon region. Hence, this mode can, in principle, be present in RNdS background. However, in our analysis, we will later impose a boundary condition that eliminates the $a_2$ mode. We will focus exclusively on the mode associated with $a_1$, which also aligns with the conventional treatment in pure de Sitter quantum field theory computations.

Now, we consider the asymptotic limit of the FHR solution near the cosmological horizon located at $r = L$. For convenience, we transform to the Tortoise coordinate $r = L\tanh(x/L)$ such that $x\to\infty$ as $r\to L$. The asymptotic form is given as,
\begin{align}
    v_{\omega l}(r(x))|_{\text{FHR}} \xrightarrow[r(x)\to L]{} a_{in}\text{e}^{-i\omega x} + a_{out}\text{e}^{+i\omega x}.
\end{align}
These solutions are in/outgoing plane waves near the cosmological horizon. The in/outgoing coefficients are given as,
\begin{align}\label{CH-amp}
    & a_{in} = \frac{ i2^{-l-1}\sqrt{\pi}}{L\sinh(\pi\omega L)\Gamma(1+i\omega L)}\left[\frac{a_2 2^{2l+1}\Gamma\left(\frac{3}{2}+l\right)}{(l+1-i\omega L)\Gamma(l-i\omega L)} - \frac{a_1 \Gamma\left(\frac{1}{2} - l\right)}{(l+i\omega L)\Gamma(-1-l-i\omega L)}\right], \\
    & a_{out} = -\frac{ i2^{-l-1}\sqrt{\pi}}{L\sinh(\pi\omega L)\Gamma(1-i\omega L)}\left[\frac{a_2 2^{2l+1}\Gamma\left(\frac{3}{2}+l\right)}{(l+1+i\omega L)\Gamma(l+i\omega L)} - \frac{a_1 \Gamma\left(\frac{1}{2} - l\right)}{(l-i\omega L)\Gamma(-1-l+i\omega L)}\right].
\end{align}
 
\subsection{Semiclassical greybody factor in static universe}
Semiclassical QFT computations done near a flat space black hole horizon predict the emission of thermal radiation. This radiation, however, has to travel through the black hole potential to reach asymptotic infinity, and doing so, it deviates from a perfect black-body spectrum. This deviation is captured by the frequency dependent `greybody factor' that multiplies the thermal distribution. For each mode, the greybody factor is computed from the transmission probability of the mode through the potential (\ref{schrodinger-bh}). 

For de Sitter black holes, the semiclassical prediction depends on the choice of vacuum for the far-horizon QFT. If we assume a static vacuum, then the semiclassical spectrum behaves analogously to that of the flat space black holes. We compute the greybody factors for this choice, expanding on the work of \cite{Harmark:2007jy}. However, later we will see that when the global de Sitter vacuum is considered, the semiclassical predictions change due to the thermal nature of the cosmological horizon.

When an ingoing beam of particles is sent from the past cosmological horizon towards the bulk, part of it falls into the black hole horizon, while the rest gets reflected to the future cosmological horizon. The ratio of the ingoing flux at the black hole horizon to the ingoing flux at the cosmological horizon gives the greybody factor. Using the conservation of total flux, it can be written as,
\begin{align}
    \gamma_l(\omega) = 1 - \frac{\abs{a_{out}}^2}{\abs{a_{in}}^2}.
\end{align}
Here, $a_{in/out}$ are the incoming and outgoing amplitudes at the cosmological horizon respectively \eqref{CH-amp}. In the small frequency limit, we find:
\begin{align}\label{dS-grey}
    & \gamma_0(\omega) \xrightarrow[\omega \to 0]{} \frac{4r_{+}^2}{L^2}(1 + \omega^2L^2 + \mathcal{O}(\omega^4)), \\
    & \gamma_l(\omega)  \xrightarrow[\omega\to 0]{} 4^{-l}L^{-2l}\pi^{2+2l}r_{+}^{2+4l}\beta^{-2l}\left(\frac{\Gamma(l)\Gamma(l+2)}{\Gamma(l+1/2)\Gamma(l+3/2)}\right)^2\omega^2 + \mathcal{O}(\omega^4).
\end{align}
The expression of $\gamma_0(\omega)$ matches that of \cite{Harmark:2007jy}. In asymptotically flat spacetime, the small frequency expansion of $\gamma_l(\omega)$ starts from $\omega^{2l+2}$. While in asymptotically static de Sitter, the small frequency expansion of $\gamma_l(\omega)$ has all the even powers $\omega^2,\cdots \omega^{2l}$ till the flat space contribution appears at an order $\omega^{2l+2}$. The static de Sitter effects become more and more important as we take smaller frequencies such that $\omega L < 1$. We are also listing the greybody factors for small values of $l$:
\begin{align}
    & \gamma_1(\omega) \xrightarrow[\omega L\to 0]{} \frac{64 \pi^2r_{+}^6\omega^2}{9\beta^2L^2}\left(1 + \frac{1}{4}\omega^2 L^2 + \cdots\right), \\
    &    \gamma_2(\omega) \xrightarrow[\omega L\to 0]{} \frac{256 \pi^4r_{+}^{10}\omega^2}{225\beta^4L^4}\left(1 + \frac{10}{9}\omega^2 L^2 + \frac{1}{9}\omega^4L^4 + \cdots\right).
\end{align}
As mentioned above, the $\omega^2,\omega^4,\omega^6$ terms in the expressions of $\gamma_0,\gamma_1,\gamma_2$ respectively match with the asymptotically flat space results \cite{Brown:2024ajk}.

As we have mentioned earlier, the semiclassical greybody spectrum fails to accurately describe the evaporation of black holes at very small temperatures. To describe these near-extremal black holes, we will now incorporate the leading quantum gravitational effects.

\section{Quantum-corrected black hole evolution}\label{sec:BH-evol}
In this section, we will discuss the evaporation rate of near-extremal de Sitter black holes using the Schwarzian effective description. As a simple toy model, we consider the evaporation through a massless, uncharged scalar field coupled to the black hole background.

\subsection{Effective picture: Schwarzian coupled to thermal bath}
As discussed in the previous section, the semiclassical description of near-extremal black holes breaks down when $\beta \gg Q$. In this regime, the quantum corrections dominate and significantly modify the thermodynamics. We will consider the effective description of the near-extremal near-horizon throat to understand the evolution of these black holes. We will adopt the ideas of \cite{Brown:2024ajk} while incorporating the thermal nature of the asymptotic far-horizon de Sitter field theory. 

First, let us briefly review the techniques of \cite{Brown:2024ajk}, as applied to the de Sitter black holes when the far horizon QFT is in static vacuum. Let us look at the falloff \eqref{BH-sol-match} of the scalar field in the overlapping region. From the 2D near-horizon throat perspective, the $r^l$ mode can be thought of as the non-normalizable solution in AdS$_2$. While the $r^{-l-1}$ mode is a normalizable fluctuation tunneling out of the near-horizon region. Due to the decoupling between the near-horizon and far-horizon regions in the small near-extremal limit, this tunneling mode does not contribute to leading order transitions. Hence we turn it off in our computations. The non-normalizable mode acts as a source at the boundary of AdS$_2$, where the Schwarzian theory lives. From \eqref{FHR-sol} and \eqref{FHR-CH-lim}, we see that the asymptotic de Sitter solutions with coefficients $a_1$ and $a_2$ approach the non-normalizable and normalizable solutions of the near-horizon throat. To turn off the tunneling mode, we set $a_2 = 0$. In asymptotically flat spacetime, \eqref{FHR-sol} is replaced by the corresponding flat space solutions that are given in terms of certain Bessel functions. The qualitative features in the overlapping region remain the same.

The main idea is to couple the near-horizon effective theory with the far-horizon QFT through the non-normalizable mode. This mode acts as a source for an operator of a particular conformal dimension in the throat. The conformal dimension is fixed by the rules of AdS$_2$/CFT$_1$. This introduces a time-dependent perturbation to the Schwarzian system. The Hamiltonian density is then given by,
\begin{align}\label{hamil}
    H = H_{\text{Sch}} + \mathcal{O}_{\Delta}(t) \left(r^{-l}\Phi_{\text{Asymp}} (t,r,\Omega)\right)_{r\to 0}.
\end{align}
Here, $\Phi_{\text{Asymp}}$ is the asymptotic field corresponding to a particular spherical harmonic. We have dropped the $l,m$ labels, and we will focus on the s-wave sector with $l = 0$, which dominates the evolution. The $r^{-l}$ factor extracts the correct non-normalizable mode. The $r\to 0$ limit of the asymptotic solution brings it to the overlapping region, where the near-horizon AdS$_2$ boundary lives. The conformal dimension $\Delta$ depends on the mass and harmonics label of the scalar field. For a massless s-wave scalar $\Delta = 1$.

The eigenstates of the $H_{\text{Sch}}$ denote different near-extremal states, with the eigenvalues denoting the mass above extremality for those black holes. The initial black hole state is $\ket{E_i}$ with energy $E_i$ while the far-horizon QFT is in the static vacuum. The interaction with the far-horizon QFT can cause a transition of the Schwarzian system to a different state $\ket{E_f}$. Due to the conservation of total energy corresponding to the timelike Killing vector $\partial_t$, the far-horizon QFT also transitions from the vacuum state to an excited state with energy $(E_f-E_i)$. The maximum contribution to the evaporation rate comes from single particle excitations. This rate is then computed using Fermi's Golden Rule. Finally, we integrate over all possible final states. The black hole evaporation rate near-extremality ($\beta\gg Q$) in the canonical ensemble is then given by,
\begin{align}\label{static-rate}
    & \frac{dE}{dt}\Big|_{\text{dS Static}} \approx \left( 1 + \frac{35\beta^2}{96L^2}\right) \frac{dE}{dt}\Big|_{\text{flat}}, \quad \frac{dE}{dt}\Big|_{\text{flat}} = \frac{256\sqrt{2E_0}\, r_{+}^2}{35\, \pi^{5/2} \beta^{7/2}}. 
\end{align}
The leading order flat space contribution is obtained in \cite{Brown:2024ajk}. In the semiclassical limit ($\beta\sim Q$), the spectrum is thermal, with a greybody factor given by \eqref{dS-grey}. The details of these calculations are given in the appendix \ref{app:static}. 

However, this calculation faces several issues in de Sitter spacetime. The static vacuum is not de Sitter invariant, and it does not capture the global features of the spacetime. Physically, the far-horizon QFT should be placed in the Bunch–Davies vacuum. But this is not a natural invariant state with respect to the static time. Hence, the Fermi Golden Rule computation does not go through as it requires the initial and final states to be eigenstates of the unperturbed Hamiltonian. Another issue is that the static model does not account for the thermal nature of the cosmological horizon. In principle, sufficiently colder black holes should be able to absorb energy from the cosmological horizon. To address these limitations, we will work with the Bunch-Davies vacuum for the far horizon QFT. This will naturally incorporate the thermal nature of the de Sitter cosmological horizon and help us understand both emission ($\beta < L$) and absorption ($\beta > L$) in the near-extreme regime when $\beta \gg r_0$. In this picture, the black hole is effectively modelled as a quantum mechanical Schwarzian system inside the thermal environment of far horizon QFT. This allows us to adopt the ideas of \cite{Higuchi:1986ww, Spradlin:2001pw}. We will use time-dependent perturbation theory to compute the transition of the Schwarzian system, while the asymptotic QFT can transition to a complete set of states. Unlike Fermi Golden Rule, this transition depends only on the coupling between the two systems, not on the detailed structure of the final state of the far-horizon QFT.

For convenience, going forward we will write $\Phi_{\text{Asymp}}\equiv\Phi$ and consider the dominant s-wave channel. We want to understand the evolution of the system at large time scales. The interaction Hamiltonian \eqref{hamil} causes a transition of the initial state $\ket{E_i}\otimes\ket{0}_{BD}$ at $t \xrightarrow{} -\infty$ to the final state $\ket{E_f}\otimes\ket{\psi}$ at $t \xrightarrow{} \infty$. Here, $\ket{E_i}, \ket{E_f}$ are states in the Schwarzian system, $\ket{0}_{BD}$ is the de Sitter invariant Bunch-Davies vacuum, and $\ket{\psi}$ is an excited state of the far-horizon QFT. In first-order perturbation theory, this particular process has a transition amplitude,
\begin{align}
    T_{if;\psi} = -i\int_{-\infty}^{+\infty} dt' \bra{E_f,\psi} \mathcal{O}_{\Delta}(t') \Phi(t',r\to 0) \ket{E_i,0}.
\end{align}
We consider the time evolution of the operator $\mathcal{O}_{\Delta} (t) = \text{e}^{i H_{\text{Sch}}t}\mathcal{O}_{\Delta}(0)\text{e}^{-i H_{\text{Sch}}t}$ with respect to the free Schwarzian Hamiltonian. This gives us,
\begin{align}
    \bra{E_f}\mathcal{O}_{\Delta}(t)\ket{E_i} = \text{e}^{i(E_f - E_i)t}\bra{E_f}\mathcal{O}_{\Delta}(0)\ket{E_i} \equiv \text{e}^{i(E_f - E_i)t} \mathcal{O}_{\Delta; fi}.
\end{align}
Now, we want to trace over the states of the environment. We impose the completeness condition of the far-horizon states,
\begin{align}
    \sum_{\psi}\ket{\psi}\bra{\psi} = 1.
\end{align}
Thus, the transition probability of the Schwarzian system going from $\ket{E_i}$ to $\ket{E_f}$ is given by,
\begin{align}\label{prob}
    \mathcal{P}_{if} \equiv &\sum_\psi T_{if;\psi}^{*} T_{if;\psi} \nonumber \\
    =& \abs{\mathcal{O}_{\Delta; fi}}^2 \int_{-\infty}^{+\infty} dt'' \int_{-\infty}^{+\infty} dt' \text{e}^{-i(E_f - E_i)(t''-t')}G(t'',r\to 0;t',r\to 0),
\end{align}
Here, $\abs{\mathcal{O}_{\Delta; fi}}^2$ is the Schwarzian correlation function \eqref{sch-corr}. $G(x;x')\equiv\bra{0}\Phi(x)\Phi(x')\ket{0}$ is the two-point function in de Sitter invariant vacuum. As discussed in section \ref{subsec:dS-qft}, the minimally coupled massless scalar Green's function in de Sitter has subtleties due to IR divergences. We circumvent this issue by regularizing the computation with a small mass and taking the massless limit in the end. As we will see, only the short-distance singular behavior of the Green's function enters our computation, and this singular behavior becomes independent of the mass. Thus, the regularized result works for the massless case. The expression for the massive Green's function in $d$-dimensions is given by \cite{Spradlin:2001pw},
\begin{align}\label{green}
    G(x;x') = \frac{\Gamma(h_{+})\Gamma(h_{-})}{(4\pi)^{d/2}\Gamma(d/2)}F\left(h_{+},h_{-};\frac{d}{2};\frac{1+\sigma(x,x')}{2}\right).
\end{align}
Here $\sigma(x,x')$ is the de Sitter invariant distance. The weights $h_{\pm}$ are given by,
\begin{align}
    h_{\pm} = \frac{1}{2}\left((d-1)\pm\sqrt{(d-1)^2 - 4m^2}\right).
\end{align}
The Green's function has poles when $\sigma(x,x') = 1$ and the singular behavior near the poles are given as,
\begin{align}\label{green-sing}
    G_{\sigma\to 1}(x;x') \sim \frac{\Gamma(d/2 -1)}{(4\pi)^{d/2}}\left(\frac{D^2(x,x')}{4}\right)^{1-d/2}.
\end{align}
Here $D(x,x') = \cos^{-1}\sigma(x,x')$. Thus, we see that the singular behavior does not depend on $h_{\pm}$ and is only a function of the invariant distance. Exploiting this feature, we will compute the probability $\mathcal{P}^{\text{reg}}_{if}$ for a massive scalar field and take the massless limit at the very end. We are interested in the `deep bulk' limit of the correlator inside the static patch. Thus, we express \eqref{green} in static coordinates and take the $r\to 0$ limit. Since the spatial separation is zero at the near horizon boundary, the invariant distance becomes $\sigma(x,x') = \cosh\left(\frac{t-t'}{L}\right)$. Thus, the correlator and the integrand only depend on the combination $(t'-t)$. Using this, we can change to variables $t_{\pm} = t''\pm t'$. Then we have,
\begin{align}
    \mathcal{P}^{\text{reg}}_{if} = \abs{\mathcal{O}_{\Delta; fi}}^2 \int_{-\infty}^{+\infty} dt_{+} \int_{-\infty}^{+\infty} dt_{-} \text{e}^{-i(E_f - E_i)t_{-}}G(t_{-}).
\end{align}
The divergent factor of the $t_{+}$ integral cancels when we compute the transition rate i.e. the transition probability per unit time. We also drop the minus subscript from $t_{-}$.
\begin{align}\label{rate-reg}
    \Gamma^{\text{reg}}_{if} \equiv \frac{\partial \mathcal{P}^{\text{reg}}_{if}}{\partial t_+} = \abs{\mathcal{O}_{\Delta; fi}}^2 \Tilde{G}(E_f - E_i).
\end{align}
Here, $\tilde G$ is the spectral density,
\begin{align}\label{spec-int}
    \tilde G(E_f - E_i) \equiv \int_{-\infty}^{+\infty} dt \text{e}^{-i(E_f - E_i)t} G(t).
\end{align}
The integrand has poles whenever $\sigma = 1$,
\begin{align}
    \sigma(t) = \cosh (t/L) = 1 \implies t = 2\pi i n L; \quad n\in \mathbb{Z}. 
\end{align}
The $n = 0$ pole is the usual flat space singularity. We also need an appropriate $i\epsilon$ prescription that shifts $t\to (t-i\epsilon)$ for $\epsilon>0$. This pushes the $t = 0$ pole upwards along the imaginary $t$-line. The singular behaviour \eqref{green-sing} of the correlation function is given by,
\begin{align}
    G(t)|_{t\to 2\pi i n L} \sim -\frac{1}{4\pi^2}\frac{1}{(t - 2\pi i n L -i \epsilon)^2}.
\end{align}
The integral \eqref{spec-int} can be computed using complex analysis. We can compute the integral along a semicircular contour consisting of the real line and a large arc that cuts the imaginary line between the $M$th and $(M+1)$th poles for large $M$. Depending on the sign of $E_f-E_i$, the arc is closed either in the upper half or in the lower half plane so that the integrand dies off along the arc. Then the contribution to the integral \eqref{spec-int} comes from the residue of the poles enclosed. Finally, an $M\to\infty$ limit is taken. 

The result of this integration is independent of the mass regulator. Naturally, we can consider a massless limit of this result. The form of the spectral density is then given by,
\begin{align}\label{spec}
     \tilde G(E_f - E_i) = \int_0^{\infty} d\omega \Big[ &\frac{\omega}{2\pi} ( N(2\pi L;\omega) + 1) \delta(\omega + E_f - E_i) \nonumber \\
     - &\frac{\omega}{2\pi} N(2\pi L;\omega) \delta(\omega - E_f + E_i) \Big].
\end{align}
Here the function $N$ is the Planck distribution factor,
\begin{align}
    N(\beta;\omega) \coloneqq \frac{1}{\text{e}^{\beta\omega} - 1}.
\end{align}
Evidently, $2\pi L$ is the inverse temperature associated with the cosmological horizon. Similar results on the spectral function were obtained in \cite{Alicki:2023rfv}. From \eqref{spec} we can note that when $E_f<E_i$, only the first delta function contributes; while for $E_f>E_i$, only the second delta function contributes. We drop the `reg' superscript henceforth. Next, we will use this result to compute the transmission rates.

\subsection{Energy transfer rate: semiclassical vs quantum}\label{subsec:rate}
We want to compute the rate of change of the energy of the near-extremal black hole at energy $E_i$. This requires calculating transitions to an arbitrary state of energy $E_f$ and then integrating over all such states with an appropriate measure. This measure comes because the Schwarzian theory has a continuous density of states \eqref{sch-dos}. Thus, we have:
\begin{align}
  \int_{0}^{\infty} dE \rho(E) \ket{E}\bra{E} = 1.
\end{align}
Using \eqref{rate-reg}, the rate of change in energy of the Schwarzian system is then given by,
\begin{align}
    \frac{dE}{dt} = \mathcal{N} \int_0^{\infty}\abs{E_f - E_i} dE_f \rho(E_f) \abs{\mathcal{O}_{\Delta; fi}}^2 \Tilde{G}(E_f - E_i).
\end{align}
We have included an appropriate normalization constant $\mathcal{N} = 4r_{+}^2$\footnote{The normalization is coming from the source term. This can be obtained by dimensional reduction of the scalar action to the throat and demanding that the lower-dimensional action is canonically normalized.}. We use the notation for the Schwarzian part,
\begin{align}\label{sch-amp}
    \mathcal{P}_{\text{Sch}}(E_i,E_f) \coloneqq \rho(E_f) \abs{\mathcal{O}_{\Delta; fi}}^2.
\end{align}
Using the expression \eqref{spec}, we perform the $E_f$ integral to obtain,
\begin{align}\label{energy-rate}
    \frac{dE}{dt} \equiv \frac{dE}{dt}^{+} + \frac{dE}{dt}^{-} = \mathcal{N}\int_0^{\infty} d\omega\ \omega \Big[  &\frac{\omega}{2\pi} ( N(2\pi L;\omega) + 1) \mathcal{P}_{\text{Sch}}(E_i,E_i - \omega) \nonumber \\
     - &\frac{\omega}{2\pi} N(2\pi L;\omega) \mathcal{P}_{\text{Sch}}(E_i,E_i + \omega)  \Big].
\end{align}
The first term describes the emission of energy $\omega$ and the second term describes absorption by the black hole. They are denoted by $\frac{dE}{dt}^{\pm}$ respectively. Since the Schwarzian density has support for non-negative energies only, the emission integral gets a contribution for $0<\omega<E_i$.

Let us first discuss the semiclassical energy transfer rate. This regime is specified by $E_i\gg E_0$ i.e. the energy of the black hole is much higher than the semiclassical physics breakdown scale. In this limit, the Schwarzian probabilities take the form,
\begin{align}
    & \mathcal{P}_{\text{Sch}}(E_i,E_i - \omega) \approx N(\beta;\omega), \nonumber \\
    & \mathcal{P}_{\text{Sch}}(E_i,E_i + \omega) \approx N(\beta;\omega) + 1.
\end{align}
Here $\beta \equiv \sqrt{\frac{2\pi^2}{E_i E_0}}$ denotes the inverse temperature of the black hole at the saddle point. Thus, the energy transfer rate \eqref{energy-rate} becomes,
\begin{align}\label{energy-rate-semicl}
    \frac{dE}{dt}\Big|_{\text{semicl}} = \mathcal{N}\int_0^{\infty} d\omega\ \frac{\omega^3}{2\pi}\left[N(\beta;\omega) - N(2\pi L;\omega)\right].
\end{align}
Here, the first term in the integrand is the flat space result of thermal distribution with the semiclassical greybody factor $\propto \omega^2$. Performing the integration, we get:
\begin{align}\label{semicl-rate}
    \frac{dE}{dt}\Big|_{\text{semicl}} = 4r_{+}^2 \frac{\pi^{3}}{30}\left( \frac{1}{\beta^{4}} -\frac{1}{\beta_C^4} \right).
\end{align}
This is just the energy transfer rate for a thermal system coupled to a thermal bath at a different temperature $\beta_C = 2\pi L$. We can naturally see emission or absorption schematically depending on $\beta < \beta_C$ and $\beta > \beta_C$, respectively. The rate also flips sign as $\beta = \beta_C$. Now we turn our attention to the quantum regime $E_i\ll E_0$, where the semiclassical physics breaks down and the Schwarzian modes become important.

In the quantum regime, the Schwarzian probabilities are given by,
\begin{align}\label{quant-sch-prob}
    & \mathcal{P}_{\text{Sch}}(E_i,E_i - \omega) \approx \frac{\sqrt{E_0/2}}{\pi}\sqrt{E_i - \omega}, \nonumber \\
    & \mathcal{P}_{\text{Sch}}(E_i,E_i + \omega) \approx \frac{\sqrt{E_0/2}}{\pi}\sqrt{E_i + \omega}.
\end{align}
The thermal factors appearing in the de Sitter spectral functions can be expanded into power series:
\begin{align}
    N(\beta_C;\omega) = \sum_{n=1}^{\infty} \text{e}^{-\beta_C\omega n}.
\end{align}
This is because for all values of $\omega\in(0,\infty)$ the exponential $\text{e}^{-\beta_C\omega n}<1$. Also, at $\omega = 0$, the full integrand is regular. Thus, we can use this series expansion of the spectral function. We further perform a change of variables $\omega\to E_i u$ in the integrals \eqref{energy-rate}. The rate of energy emission is then given by,
\begin{align}
    \frac{dE}{dt}^{+} = \frac{r_{+}^2\sqrt{2E_0}}{\pi^2}E_i^{7/2}\sum_{n=0}^{\infty}\int_0^1 u^2\sqrt{1-u}\ \text{e}^{-\beta_C E_i n u} du
\end{align}
The integral appearing here is the integral representation of the Kummer function of the first kind \eqref{kummer-func}. Using \eqref{kummer-int}, it can be written as,
\begin{align}
    \frac{dE}{dt}^{+} = \frac{r_{+}^2\sqrt{2E_0}}{\pi^2}E_i^{7/2}\frac{\Gamma(a)\Gamma(b-a)}{\Gamma(b)}\sum_{n=0}^{\infty} M(a,b,-z_n),
\end{align}
Here $a = 3, b = 9/2$ and $z_n = \beta_C E_i n$. Similarly, we can perform the same change of variables and \eqref{kummer-int} to express the absorption rate in terms of Kummer function of the second kind:
\begin{align}
    \frac{dE}{dt}^{-} &= -\frac{r_{+}^2\sqrt{2E_0}}{\pi^2}E_i^{7/2}\sum_{n=1}^{\infty}\int_0^{\infty} u^2\sqrt{1+u}\ \text{e}^{-\beta_C E_i n u} du \nonumber \\
    &= -\frac{r_{+}^2\sqrt{2E_0}}{\pi^2}E_i^{7/2}\Gamma(a)\sum_{n=1}^{\infty}U(a,b,z_n).
\end{align}
The full energy transfer rate is given as,
\begin{align}\label{quantum-rate}
    \boxed{\frac{dE}{dt} = \frac{r_{+}^2\sqrt{2E_0}}{\pi^2}E_i^{7/2}\left[ \frac{16}{105} + \sum_{n=1}^{\infty}\left( \frac{16}{105}M(a,b,-z_n) - 2U(a,b,z_n) \right) \right]}
\end{align}
The first term here is the flat space contribution, and it matches with the results of \cite{Brown:2024ajk}. The terms in the summation are the de Sitter corrections. This is one of our main results. 

We can also investigate subleading corrections to this result by considering the next-order behaviour of the Schwarzian probabilities in the quantum regime \eqref{quant-sch-prob}. Including such corrections gives us a contribution to the rate,
\begin{align}
    \frac{dE}{dt}^{\text{subleading}} = -\frac{2\sqrt{2}r_{+}^2}{3\sqrt{E_0}}\left[\int_0^{E_i}d\omega\frac{\omega^3\sqrt{E_i-\omega}}{1-\text{e}^{-\beta_C \omega}} + \int_0^{\infty}d\omega\frac{\omega^3\sqrt{E_i+\omega}}{\text{e}^{\beta_C \omega}-1} \right]
\end{align}
The integrations can be manipulated similarly, and the final result is as follows,
\begin{align}\label{quant-sublead}
    \frac{dE}{dt}^{\text{subleading}} = -\frac{2\sqrt{2}r_{+}^2}{3\sqrt{E_0}} E_i^{9/2}\left[ \frac{32}{315} + \sum_{n=1}^{\infty}\left( \frac{32}{315}M(a',b',-z_n) + 6U(a',b',z_n) \right) \right],
\end{align}
Here $a'=4,b'=11/2$. As expected, these are much suppressed.

\subsection{Hotter than cosmos vs colder than cosmos}

\begin{figure}[htbp]
    \centering
    \includegraphics[width=0.7\textwidth]{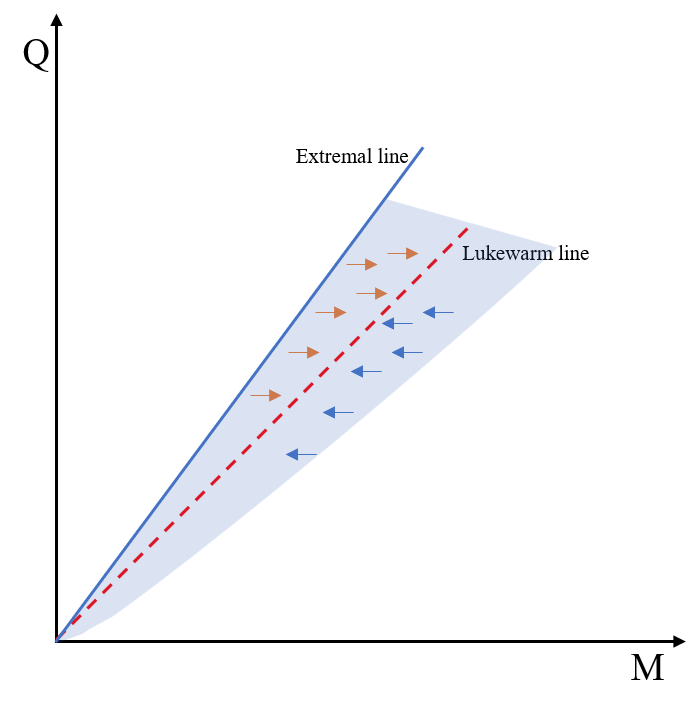}
    \caption{Evolution of small near-extremal black holes. Here we zoomed in near the cold branch for small black holes in phase space.}
    \label{fig:small-NE}
\end{figure}

While it is difficult to understand the physical behavior of the de Sitter corrections directly from this expression, we can gain more insight by analyzing its asymptotic behaviors on either side of the lukewarm line. This line is the thermal equilibrium curve i.e. along this line, the black hole and cosmological horizon temperatures are equal. From the thermodynamic relations, we find that the energy of a small lukewarm near-extremal black hole satisfies,
\begin{align}
    E_i^{\text{lukewarm}} L \sim \frac{r_0^3}{L^2}L.
\end{align}
As we are considering a small black hole satisfying $r_0\ll L$, we find that in the quantum regime $E_i\ll E_0$, the lukewarm black holes have very small energies such that $E_i^{\text{lukewarm}} L \leq 1$. Therefore, small near-extremal black holes satisfying ${E_i L \gg 1; E_i\ll E_0}$ lie on the hotter side of the lukewarm line. The opposite limit i.e. $E_i L \ll 1$ is more subtle because it's not apriori evident whether these black holes would lie on the colder side of the lukewarm line. We can still estimate the leading $E_i L \to 0$ behaviour which is very close to the extremal line and this behaviour would explain how an extremal black hole will absorb energy. A cartoon of these two processes is depicted in figure \ref{fig:small-NE}. We explore these two limits next.

\subsubsection*{Hotter than cosmos}
Firstly, we consider the hotter than cosmos limit $E_iL\gg 1$. Since in the de Sitter corrections of \eqref{quantum-rate} $n>1$, we have $z_n = \beta_C E_i n \gg 1$. In this limit, we can use the large-$z_n$ behavior \eqref{large-z} of the Kummer functions. As $a=3, b=9/2$, the required conditions for the expansions are met. We consider the $n$-th summand of the de Sitter correction,
\begin{align}
    &\frac{16}{105}M(a,b,-z_n) - 2U(a,b,z_n) \nonumber \\
    =& \frac{16}{105}\frac{\Gamma(b) (-1)(-z_n)^{-a}}{\Gamma(b-a)} \sum_{k=0}^{\infty}\frac{(a)_k (a-b+1)_k}{k!}(+z_n)^{-k} 
    -2 z_n^{-a}  \sum_{k=0}^{\infty}\frac{(a)_k (a-b+1)_k}{k!}(-z_n)^{-k} \nonumber \\
    =& 2 z_n^{-a}  \sum_{k=0}^{\infty}\frac{(a)_k (a-b+1)_k}{k!}z_n^{-k}\left[1 - (-1)^k \right] \nonumber \\
    =& 4 z_n^{-a}  \sum_{k=1,3,5\cdots}^{\infty}\frac{(a)_k (a-b+1)_k}{k!}z_n^{-k}
\end{align}
Now, in the $z_n\gg 1$ limit, the leading behavior comes from the $k=1$ term. This gives,
\begin{align}\label{dS-large-zn}
    \frac{16}{105}M(a,b,-z_n) - 2U(a,b,z_n) \sim 4a(a-b+1) z_n^{-a-1} = -\frac{6}{(\beta_C E_i n)^4}.
\end{align}
In the full de Sitter correction, we also have to perform a sum over $n$, which results in a zeta function \eqref{zeta}. This gives,
\begin{align}\label{quant-hot}
    \frac{dE}{dt}^{\text{hotter}} \sim \frac{r_{+}^2\sqrt{2E_0}}{\pi^2}E_i^{7/2}\left[ \frac{16}{105} - \frac{6\zeta(4)}{\beta_C^4 E_i^4} \right].
\end{align}
Naturally, only the hotter side of the lukewarm line has an analog in the flat space. Thus, we consider the ratio of rates in de Sitter and flat space,
\begin{align}
    \frac{\dot{E}^{\text{hotter}}}{\dot{E}^{\text{flat}}} \sim 1 - \frac{7\pi^2}{16\beta_C^4 E_i^4}.
\end{align}
The ratio being less than unity is a consequence of the fact that, unlike flat space, the black hole in de Sitter radiates into a thermal background. The presence of this ambient radiation essentially suppresses the net emission rate. 

We can similarly estimate the subleading corrections \eqref{quant-sublead} in the hotter regime. This is given by,
\begin{align}
    \frac{dE}{dt}^{\text{hotter,subleading}} \sim -\frac{2\sqrt{2}r_{+}^2}{3\sqrt{E_0}}E_i^{9/2}\left[ \frac{32}{315} - \frac{12\zeta(5)}{\beta_C^5 E_i^5} \right].
\end{align}

\subsubsection*{Colder than cosmos}
Next, we consider the opposite limit $E_iL\ll 1$ and extract the leading $E_iL\to 0$ behaviour. Unlike flat space, the extremal black holes in de Sitter are not at thermal equilibrium. Thus these black holes should absorb energy from the environment to reach the lukewarm line. Given such an energy $E_i$, we can find a large integer $M\sim\mathcal{O}\left(\frac{1}{E_iL}\right)$ such that:
\begin{align}
    z_n<1; \quad n<M\quad \text{and} \quad z_n>1; \quad n\geq M.
\end{align}
Hence, we need to split the sum into two parts in the de Sitter correction of \eqref{quantum-rate}. In the first part of the sum over $1\leq n < M$, the leading asymptotic behavior is governed by the small-$z_n$ approximation of Kummer functions. Whereas, in the second part of the sum $M\leq n < \infty$, the leading behavior can be given by the large-$z_n$ asymptotics of the Kummer functions. 

Let us consider the $n$-th summand of \eqref{quantum-rate} for $z_n\ll 1$. We use the series representation \eqref{small-z} and extract the leading behaviour,
\begin{align}
    &\frac{16}{105}M(a,b,-z_n) - 2U(a,b,z_n) \nonumber \\
    \sim&  \frac{16}{105} - \frac{2\Gamma(1-b)}{\Gamma(a-b+1)} - \frac{2\Gamma(b-1)}{\Gamma(a)}z_n^{1-b} \nonumber \\
    =& \frac{32}{105} - \frac{15\sqrt{\pi}}{8}z_n^{-7/2}
\end{align}
Computing the sum over $1\leq n\leq M-1$, we get the de Sitter contribution from small $z_n$ regime,
\begin{align}
    \frac{32}{105}(M-1) - \frac{15\sqrt{\pi}}{8}(\beta_C E_i)^{-7/2}H_{M-1,7/2}.
\end{align}
$H_{m,k}$ is the generalized harmonic number \eqref{harm-num}. In the large-$M$ limit, the above expression takes the form,
\begin{align}\label{cold-small}
    \frac{32}{105}M - \frac{15\sqrt{\pi}}{8}(\beta_C E_i)^{-7/2}\left(\zeta(7/2)-\frac{2}{5}M^{-5/2}\right).
\end{align}

We can estimate the contribution from the sum over $M\leq n\leq \infty$ using the computations of the hotter black holes \eqref{dS-large-zn}. This is given by,
\begin{align}
    -\frac{6}{\beta_C^4 E_i^4}\sum_{n=M}^{\infty}\frac{1}{n^4} = -\frac{\psi^{(3)}(M)}{\beta_C^4 E_i^4}.
\end{align}
$\psi^{(m)}(z)$ is the Polygamma function of order $m$ \eqref{polygamma}. In the large-$M$, this contribution takes the form,
\begin{align}\label{cold-large}
    -\frac{2M^{-3}}{\beta_C^4 E_i^4}.
\end{align}
Substituting the asymptotic forms \eqref{cold-small} and \eqref{cold-large} into the quantum rate \eqref{quantum-rate}, we obtain,
\begin{align}
    \frac{dE}{dt}^{\text{colder}} \sim \frac{r_{+}^2\sqrt{2E_0}}{\pi^2}E_i^{7/2}\left[ \frac{16}{105} + \frac{32}{105}M - \frac{15\sqrt{\pi}}{8}(\beta_C E_i)^{-7/2}\left(\zeta(7/2)-\frac{2}{5}M^{-5/2}\right) -\frac{2M^{-3}}{\beta_C^4 E_i^4}\right].
\end{align}
Since the integer is chosen so that $M\sim\mathcal{O}\left(\frac{1}{E_iL}\right)$, we can immediately read off the leading asymptotic behaviour,
\begin{align}\label{quant-cold}
    \frac{dE}{dt}^{\text{colder}} \sim -\frac{r_{+}^2\sqrt{2E_0}}{\pi^2} \frac{15\sqrt{\pi}}{8}\beta_C^{-7/2}\zeta(7/2). 
\end{align}
This shows that the rate is negative, which indicates absorption. This is in accordance with the expectation that extremal black holes in de Sitter should absorb energy. Furthermore, in the strict $L\to\infty$ limit, the rate goes to zero, indicating that the process has no analogue in flat space.

In the colder limit, the dominant behaviour of the subleading contributions \eqref{quant-sublead} takes a similar form,
\begin{align}
    \frac{dE}{dt}^{\text{colder,subleading}} \sim -\frac{2\sqrt{2}r_{+}^2}{3\sqrt{E_0}}\frac{105\sqrt{\pi}}{16}\beta_C^{-9/2}\zeta(9/2). 
\end{align}

\section{Discussions}\label{sec:disc}

In this article, we have discussed the quantum evolution of charged near-extremal black holes in de Sitter. We considered black holes of sizes much smaller compared to the characteristic de Sitter length scale. Exploiting the effective Schwarzian description in the near-horizon throat of near-extremal black holes, we compute the quantum evaporation rate of such black holes. The methodology closely follows similar works \cite{Brown:2024ajk, Bai:2023hpd} for asymptotically flat black holes but there are various subtleties that are unique to de Sitter. The charged de Sitter black hole phase space has a richer structure, as there are three horizons. Consequently, there are three kinds of extremal limits. We only zoom near the \textit{cold extremal} black holes, which have a natural analogue in asymptotically flat spacetimes. Their near-horizon geometry also has an AdS$_2$ factor similar to flat black holes. Another subtlety appears due to the thermal nature of the de Sitter cosmological horizon. Unlike flat space, the extremal black holes in de Sitter are not the equilibrium configuration. In flat space, near-extremal black holes move towards extremality, while in de Sitter, near-extremal black holes move towards lukewarm line. The lukewarm black holes have same temparture as the cosmological horizon. Thus even in the near-extremal phase space, we have two different small temperature regimes depending of which side of the lukewarm line is in consideration. \textit{Hotter} black holes emit energy while \textit{colder} black holes absorb energy from the cosmological horizon to approach a thermal equilibrium. 

 The `near-extreme' condition ensures that the geometry contains a large near-horizon throat while the `smallness' ensures that the far-horizon QFT behaves like that of empty de Sitter. The quantum theories in these two regions are coupled in an intermediate region, where the boundary effective theory resides. This coupling of the systems introduces a time-dependent perturbation to the effective Hamiltonian. This perturbation causes a transition of the black hole. We computed these transition amplitudes from the time dependent perturbation theory assuming that the far horizon QFT is initially in Bunch-Davies vacuum state. But this choice of vacuum state is not unique as we discuss below.

An analogous computation holds when the far-horizon de Sitter QFT is in the static vacuum, as illustrated in \ref{app:static}. However, this does not take into account the thermal nature of the ambient de Sitter spacetime. Also, the static vacuum is not de Sitter invariant. Thus, we should put the far-horizon QFT in a de Sitter invariant vacuum. To understand the evolution while incorporating the thermal nature of de Sitter, we have considered a toy model where the evolution is governed by minimally coupled massless scalar particles. We further treat the Schwarzian theory as an open system coupled to the de Sitter environment. We have only specified that the environment transitions into a complete set of states at the expense of the transition of the black hole to a different energy. This introduces the scalar two-point function in the de Sitter invariant vacuum to the transition rate. Minimally coupled massless scalar two-point function has infrared divergences. Thus, we have regulated our problem by introducing a small mass. Using the massive two-point function we have computed the final energy transfer rate. It turns out that the final result is independent of the mass regulator, and it also has a smooth flat space limit. In the end, we have integrated over the final black hole state.

Interestingly, our energy transfer rate has two pieces describing both emission and absorption. First, we have computed the semiclassical evolution rate by taking the asymptotic limit $E_i\gg E_0$, where $E_i$ is the energy above extremality and $E_0$ is the semiclassical physics breakdown scale. In the static vacuum, this evolution is governed by a thermal spectrum with an appropriate greybody factor that matches with existing literature \cite{Harmark:2007jy}. In contrast, in the global vacuum, the evolution is given by the spectrum of a thermal body radiating into a thermal bath at a different temperature \eqref{semicl-rate}. The global vacuum result shows both emission and absorption, depending on whether the black hole is hotter or colder than the cosmological horizon. 

Next, we have computed the quantum evolution rate with the limit $E_i \ll E_0$. The global vacuum result is given in terms of Kummer functions \eqref{quantum-rate}. For better analytical understanding, we have considered two asymptotic regimes $E_iL\gg 1$ and $E_i\ll 1$. The former corresponds to black holes that are hotter than the cosmological horizon. These black holes emit energy at a rate given in \eqref{quant-hot}. This result has a direct analogue in flat space. We find that both the static rate \eqref{static-rate} and the global rate \eqref{quant-hot} boil down to the flat space results of \cite{Brown:2024ajk} in the large $L$ limit. But there are interesting qualitative differences in the choice of vacuum. The static rate is higher than the flat space result, while the global rate is lower than that of flat space. This is because, when the black hole is embedded in the static patch, the effective potential just outside the near-horizon region is lower than that in asymptotically flat space. This causes an increase in the emission rate. Whereas the computation in the global vacuum considers the thermal nature of the cosmological horizon. This implies that the environment has positive energy, which reduces the emission rate. This feature is also present in the semiclassical results.

When we consider the outside QFT in static vacua, there cannot be absorption and the black hole will always lose energy via radiation. This is analogous to the case for flat space black holes. But in the global de Sitter vacuum, we can probe small enough energies where we can see absorption. This is computed from the $E_iL\ll 1$ regime. The leading $E_i L\to 0$ behavior illustrates how a cold extremal black hole absorbs energy. This is shown in \eqref{quant-cold} which shows that the leading behavior of the rate is a constant negative number. The negative sign indicates absorption, and it also vanishes at $L\to \infty$, consistent with the fact that there is no absorption in flat space. These findings meet the basic physical expectations. However, the result in this regime may be significantly modified due to non-perturbative effects. Also, the applicability of the effective Schwarzian picture at such low temperatures is questionable \cite{Banerjee:2023gll}.

\subsection{Effects of gauge modes}
So far, we have considered the Schwarzian modes corresponding to large diffeomorphisms in AdS$_2$. But the charged black holes in de Sitter have extra gauge modes due to the breaking of large SO(3) and U(1) gauge symmetries \cite{Blacker:2025zca}. The large SO(3) modes correspond to the enhancement of the spherical isometry of the background; they are also present in flat space \cite{Banerjee:2023quv}. The U(1) modes are a novel feature in de Sitter, and they originate from the gauge field under which the black hole is charged. To understand their contributions, the effective boundary theory should be coupled to additional gauge mode action.  The U(1) modes have an associated electric susceptibility that can be computed from the thermodynamics,
\begin{align}
    K_e = \frac{\partial Q}{\partial \mu}\Big|_{\beta\to\infty} = -\frac{L^2}{3r_0}.
\end{align}
This susceptibility parameter acts as a coupling constant in the effective theory. In the flat space limit, $K_e\to\infty$, hence the modes decouple from the Schwarzian part and do not affect the low temperature physics. But in de Sitter, they do contribute.

The correlation functions of Schwarzian theory coupled to gauge modes are studied in the literature \cite{Mertens:2018fds}. A comprehensive analysis of the contribution of the SO(3) modes is beyond the scope of our current work; we hope to revisit this in the future. But the contribution of U(1) mode does not qualitatively modify our result. This only shifts the energy by a very small parameter $E_i\to E_i - \frac{Q^2}{2K_e}$ in the Schwarzian amplitude \eqref{sch-amp}. Effects of U(1) modes were studied in the context of asymptotically flat rotating black holes \cite{Maulik:2025hax}. 

\subsection{Future directions}
Let us conclude the paper with some open problems. The natural next order question is to compute the quantum evaporation rates for Standard Model particles for both charged and rotating black holes in de Sitter. We hope to report on this. We would also like to carefully investigate the effects of the gauge modes in the transfer of charged and spinning particles in respective scenarios. It would be interesting to explore the evolution of near-Nariai black holes. These black holes also have an infinite number of symmetries coming from their dS$_2$ near-horizon throat \cite{Blacker:2025zca}. But the effects of these symmetries are less understood and require extensive analysis. Another interesting problem would be to understand the evaporation of near-extremal AdS black holes using similar techniques, and also from the boundary CFT perspective.

\acknowledgments

We are extremely grateful to Ashoke Sen for discussions and helpful insights. We are also thankful to Nabamita Banerjee, Suvrat Raju, and Tomasz R. Taylor for valuable discussions. This project was supported by the Polish National Agency for Academic Exchange under the NAWA Chair programme. MS would like to acknowledge the hospitality of the Quantum Mathematical Physics group at the University of Warsaw during the final stage of the project. Part of the results were presented at the University of Warsaw String Journal Club. Finally, we would like to thank the people of India for their generous support towards research in fundamental sciences.

\appendix

\section{Schwarzian computation with static dS vacuum}\label{app:static}
In this section, we show the computation of the quantum transfer rate with the far-horizon QFT static vacuum. This also serves as a review of the computations of \cite{Brown:2024ajk}. Let us begin with the effective Hamiltonian \eqref{hamil}. We expand the far-horizon quantum field in terms of the modes that have positive and negative frequencies with respect to static time \cite{Higuchi:1986ww}. 
\begin{align}
     &\Phi(x) = \sum_{lm}\int_0^\infty d\omega (\mathcal{V}^S_{\omega l m}(x) \hat{a}_{\omega l m} + \mathcal{V}^{S*}_{\omega l m}(x) \hat{a}^{\dagger}_{\omega l m}), 
\end{align}
Here, the mode functions are given as,
\begin{align}
    & \mathcal{V}^S_{\omega l m} (x) = M_{\omega l}\text{e}^{-i\omega t}Y_{lm}(\Omega) L^{-l}r^{l} \left(1 - \frac{r^2}{L^2}\right)^{-iL\omega/2} F\left(\frac{l-iL\omega}{2},\frac{3+l-iL\omega}{2};\frac{3}{2}+l;\frac{r^2}{L^2}\right), \nonumber \\
    & M_{\omega l} \equiv \frac{L^{-1/2}\sqrt{\sinh(\pi\omega L)}}{2\pi\Gamma(l+3/2)}\abs{\Gamma\left(\frac{3 + l - i\omega L}{2}\right)\Gamma\left(\frac{l-i\omega L}{2}\right)}.
\end{align}
These functions form an orthonormal set with respect to the standard Klein–Gordon inner product, and their coefficients appearing in the field expansion correspond to canonically quantized oscillator modes. These oscillators define the static vacuum state:
\begin{align}
    \hat{a}_{\omega l m}\ket{0}_S = 0.
\end{align}
The far-Hamiltonian is given as,
\begin{align}
     H_{\text{Static}} = \sum_{lm}\int_0^{\infty} \omega \hat a^\dagger_{\omega lm} \hat a_{\omega lm} d\omega.
\end{align}
We assume that the far-horizon QFT is initially in the static vacuum while the Schwarzian system is in a state $\ket{E_i}$. The interaction between the systems causes a transition; the black hole emits energy while the far-horizon system goes to an excited state. These transition rates are dominated by s-wave single-particle emissions. Thus, we consider the transition: $\ket{E_i}\otimes\ket{0}_S\to \ket{E_f}\otimes \hat a^\dagger_\omega\ket{0}_S$. Due to energy conservation, we also have $E_i - E_f = \omega$. 

In the interaction Hamiltonian \eqref{hamil}, we use the above mode expansion for the source term for the s-wave\footnote{We have dropped the labels $l,m$ for s-wave.},
\begin{align}
    \hat{\phi}_0(t) \equiv \left(r^{-l}\Phi_{\text{Asymp}} (x)\right)_{r\to 0} = \int_0^\infty d\omega M_{\omega} (\text{e}^{-i\omega t} \hat{a}_{\omega} + \text{e}^{i\omega t}\hat{a}^{\dagger}_{\omega}).
\end{align}
Then the transition rate is given according to Fermi's Golden rule,
\begin{align}
    \Gamma_{i\to f} = 2\pi \abs{\bra{0_S}\hat{a}_{\omega}\hat{\phi}_0\ket{0_{S}}}^2 \abs{\bra{E_f}{O}_{\Delta}\ket{E_i}}^2\delta(E_i -E_f - \omega).
\end{align}
Note that for this step, it is crucial that the initial and final states are eigenstates of the full unperturbed Hamiltonian $H_{\text{Sch}} + H_{\text{Static}}$. Integrating over the final states, the energy transfer rate is given by,
\begin{align}
    \frac{dE}{dt} = \int \omega\ d\omega \int dE_f\ \rho_{\text{Sch}}(E_f) \Gamma_{i\to f}
\end{align}
The Schwarzian computation remains the same as shown in the subsection \ref{subsec:rate}. The de Sitter contribution is modified as we are using the static vacuum. Using the properties of the oscillator modes, we find,
\begin{align}\label{static-dS-prob}
    \mathcal{P}_{dS} &= \abs{\bra{0_{S}}\hat{a}_{\omega}\hat{\phi}_0\ket{0_{S}}}^2 =\abs{\int_0^{\infty}dp M_{p}\bra{0_{S}}\hat{a}_{\omega}(\hat{a}_{p}+\hat{a}_{p}^{\dagger})\ket{0_{S}}}^2 =\frac{1 + \omega^2L^2}{\pi\omega L^2};
\end{align}
In the semiclassical limit $E_i\gg E_0$, the Schwarzian contribution gives a thermal distribution at inverse temperature $\beta$ of the black hole. The de Sitter probability obtained above exactly gives the greybody factor\footnote{We have incorporated an appropriate normalization constant into the source.},
\begin{align}
    \gamma_0(\omega) = \frac{4r_{+}^2}{L^2}(1 + \omega^2L^2).
\end{align}
This matches with \eqref{dS-grey}. In the quantum limit $E_i \ll E_0$, the Schwarzian probability is approximated as shown in \ref{subsec:rate}. After performing the integrations with the static dS contribution \eqref{static-dS-prob}, the final rate becomes,
\begin{align}
    &\frac{dE}{dt}\Big|_{\text{dS, Static}} = \frac{dE}{dt}\Big|_{\text{flat}} + \frac{2\sqrt{2E_0}r_{+}^2  E_i^{3/2}}{3 \pi^2L^2}, \quad \frac{dE}{dt}\Big|_{\text{flat}} = \frac{16\sqrt{2E_0}\, r_{+}^2 E_i^{7/2} }{105 \pi^2}.
\end{align}
The result shown in \eqref{static-rate} can be obtained from a change of ensemble from this microcanonical result.

\section{Useful identities}\label{app:id}
Here, we list the important mathematical identities used in the quantum transfer rate computation (Resources: \cite{Olver:2023DLMF, Gradshteyn:2014Ryzhik}).  \\

\noindent Kummer's equation or confluent hypergeometric equation:
\begin{align}
    z w'' + (b-z)w' - aw = 0.
\end{align}
Solutions:
\begin{align}\label{kummer-func}
    & M(a,b,z) = {}_1F_1(a; b; z), \\
    & U(a,b,z) = \frac{\Gamma(1-b)}{\Gamma(a-b+1)} M(a,b,z) + \frac{\Gamma(b-1)}{\Gamma(a)}z^{1-b}M(a-b+1,2-b,z).
\end{align}
Integral representations:
\begin{align}\label{kummer-int}
    & M(a,b,z) = \frac{\Gamma(b)}{\Gamma(a)\Gamma(b-a)}\int_0^1  \text{e}^{zu}u^{a-1}(1-u)^{b-a-1} du; \quad [\Re b > \Re a > 0], \\
    & U(a,b,z) = \frac{1}{\Gamma(a)}\int_0^{\infty} \text{e}^{-zu}u^{a-1}(1+u)^{b-a-1} du; \quad [\Re a>0, \Re z>0].
\end{align}
Series representation ($b\neq 0,-1,-2,\cdots$):
\begin{align}\label{small-z}
    M(a,b,z) = \sum_{k=0}^{\infty} \frac{(a)_k}{(b)_k k!}z^{k}; \quad (a)_k = a(a+1)\cdots(a+k-1).
\end{align}
Large $z$-expansions:
\begin{align}\label{large-z}
    & M(a,b,z) \sim \frac{\Gamma(b)\text{e}^{z}z^{a-b}}{\Gamma(a)}\sum_{k=0}^{\infty}\frac{(1-a)_k (b-a)_k}{k!}z^{-k} + \frac{\Gamma(b) \text{e}^{\pm i\pi a}z^{-a}}{\Gamma(b-a)}\sum_{k=0}^{\infty}\frac{(a)_k (a-b+1)_k}{k!}(-z)^{-k}; \nonumber \\
    &\quad \left[-\frac{\pi}{2}< \pm \arg z < \frac{3\pi}{2}; a,b,b-a\neq 0,-1,-2,\cdots\right], \\
    & M(a,b,z) \sim z^{-a} \sum_{k=0}^{\infty}\frac{(a)_k (a-b+1)_k}{k!}(-z)^{-k}; \qquad \left[-\frac{3\pi}{2}< \arg z < \frac{3\pi}{2}\right]. 
\end{align}
Zeta function:
\begin{align}\label{zeta}
    \zeta(s) = \sum_{n=1}^{\infty} \frac{1}{n^s}, \quad \Re s>1.
\end{align}
Generalized harmonic number:
\begin{align}\label{harm-num}
    H_{m,k} = \sum_{n=1}^{m} \frac{1}{n^k}, \quad \lim_{m\to\infty} H_{m,k} = \zeta(k).
\end{align}
Polygamma function:
\begin{align}\label{polygamma}
    \psi^{(m)}(z) = \frac{d^{m+1}}{dz^{m+1}}\ln\Gamma(z).
\end{align}

\bibliography{RNdS}

\providecommand{\href}[2]{#2}\begingroup\raggedright\begin{thebibliography}{10}

\bibitem{Sen:2007qy}
A.~Sen, {\it {Black Hole Entropy Function, Attractors and Precision Counting of
  Microstates}},  {\em Gen. Rel. Grav.} {\bf 40} (2008) 2249--2431,
  [\href{http://arxiv.org/abs/0708.1270}{{\tt arXiv:0708.1270}}].

\bibitem{Sen:2008vm}
A.~Sen, {\it {Quantum Entropy Function from AdS(2)/CFT(1) Correspondence}},
  {\em Int. J. Mod. Phys. A} {\bf 24} (2009) 4225--4244,
  [\href{http://arxiv.org/abs/0809.3304}{{\tt arXiv:0809.3304}}].

\bibitem{Preskill:1991tb}
J.~Preskill, P.~Schwarz, A.~D. Shapere, S.~Trivedi, and F.~Wilczek, {\it
  {Limitations on the statistical description of black holes}},  {\em Mod.
  Phys. Lett. A} {\bf 6} (1991) 2353--2362.

\bibitem{Nayak:2018qej}
P.~Nayak, A.~Shukla, R.~M. Soni, S.~P. Trivedi, and V.~Vishal, {\it {On the
  Dynamics of Near-Extremal Black Holes}},  {\em JHEP} {\bf 09} (2018) 048,
  [\href{http://arxiv.org/abs/1802.09547}{{\tt arXiv:1802.09547}}].

\bibitem{Moitra:2019bub}
U.~Moitra, S.~K. Sake, S.~P. Trivedi, and V.~Vishal, {\it {Jackiw-Teitelboim
  Gravity and Rotating Black Holes}},  {\em JHEP} {\bf 11} (2019) 047,
  [\href{http://arxiv.org/abs/1905.10378}{{\tt arXiv:1905.10378}}].

\bibitem{Iliesiu:2020qvm}
L.~V. Iliesiu and G.~J. Turiaci, {\it {The statistical mechanics of
  near-extremal black holes}},  {\em JHEP} {\bf 05} (2021) 145,
  [\href{http://arxiv.org/abs/2003.02860}{{\tt arXiv:2003.02860}}].

\bibitem{Heydeman:2020hhw}
M.~Heydeman, L.~V. Iliesiu, G.~J. Turiaci, and W.~Zhao, {\it {The statistical
  mechanics of near-BPS black holes}},  {\em J. Phys. A} {\bf 55} (2022), no.~1
  014004, [\href{http://arxiv.org/abs/2011.01953}{{\tt arXiv:2011.01953}}].

\bibitem{Sachdev:2019bjn}
S.~Sachdev, {\it {Universal low temperature theory of charged black holes with
  AdS$_2$ horizons}},  {\em J. Math. Phys.} {\bf 60} (2019), no.~5 052303,
  [\href{http://arxiv.org/abs/1902.04078}{{\tt arXiv:1902.04078}}].

\bibitem{Larsen:2018iou}
F.~Larsen, {\it {A nAttractor mechanism for nAdS$_{2}$/nCFT$_{1}$ holography}},
   {\em JHEP} {\bf 04} (2019) 055, [\href{http://arxiv.org/abs/1806.06330}{{\tt
  arXiv:1806.06330}}].

\bibitem{Castro:2018ffi}
A.~Castro, F.~Larsen, and I.~Papadimitriou, {\it {5D rotating black holes and
  the nAdS$_{2}$/nCFT$_{1}$ correspondence}},  {\em JHEP} {\bf 10} (2018) 042,
  [\href{http://arxiv.org/abs/1807.06988}{{\tt arXiv:1807.06988}}].

\bibitem{Iliesiu:2022onk}
L.~V. Iliesiu, S.~Murthy, and G.~J. Turiaci, {\it {Revisiting the Logarithmic
  Corrections to the Black Hole Entropy}},
  \href{http://arxiv.org/abs/2209.13608}{{\tt arXiv:2209.13608}}.

\bibitem{Banerjee:2023quv}
N.~Banerjee and M.~Saha, {\it {Revisiting leading quantum corrections to near
  extremal black hole thermodynamics}},  {\em JHEP} {\bf 07} (2023) 010,
  [\href{http://arxiv.org/abs/2303.12415}{{\tt arXiv:2303.12415}}].

\bibitem{Banerjee:2023gll}
N.~Banerjee, M.~Saha, and S.~Srinivasan, {\it {Logarithmic corrections for
  near-extremal black holes}},  {\em JHEP} {\bf 2024} (2024) 077,
  [\href{http://arxiv.org/abs/2311.09595}{{\tt arXiv:2311.09595}}].

\bibitem{Kapec:2023ruw}
D.~Kapec, A.~Sheta, A.~Strominger, and C.~Toldo, {\it {Logarithmic Corrections
  to Kerr Thermodynamics}},  {\em Phys. Rev. Lett.} {\bf 133} (2024), no.~2
  021601, [\href{http://arxiv.org/abs/2310.00848}{{\tt arXiv:2310.00848}}].

\bibitem{Rakic:2023vhv}
I.~Rakic, M.~Rangamani, and G.~J. Turiaci, {\it {Thermodynamics of the
  near-extremal Kerr spacetime}},  {\em JHEP} {\bf 06} (2024) 011,
  [\href{http://arxiv.org/abs/2310.04532}{{\tt arXiv:2310.04532}}].

\bibitem{Maulik:2024dwq}
S.~Maulik, L.~A. Pando~Zayas, A.~Ray, and J.~Zhang, {\it {Universality in
  logarithmic temperature corrections to near-extremal rotating black hole
  thermodynamics in various dimensions}},  {\em JHEP} {\bf 06} (2024) 034,
  [\href{http://arxiv.org/abs/2401.16507}{{\tt arXiv:2401.16507}}].

\bibitem{Modak:2025gvp}
A.~Modak, A.~Singh, and B.~Panda, {\it {Logarithmic Corrections for
  Near-extremal Kerr-Newman Black Holes}},
  \href{http://arxiv.org/abs/2502.18173}{{\tt arXiv:2502.18173}}.

\bibitem{Brown:2024ajk}
A.~R. Brown, L.~V. Iliesiu, G.~Penington, and M.~Usatyuk, {\it {The evaporation
  of charged black holes}},  \href{http://arxiv.org/abs/2411.03447}{{\tt
  arXiv:2411.03447}}.

\bibitem{Bai:2023hpd}
Y.~Bai and M.~Korwar, {\it {Near-extremal charged black holes: greybody factors
  and evolution}},  {\em JHEP} {\bf 03} (2023) 151,
  [\href{http://arxiv.org/abs/2301.07739}{{\tt arXiv:2301.07739}}].

\bibitem{Maulik:2025hax}
S.~Maulik, X.~Meng, and L.~A. Pando~Zayas, {\it {Quantum-Corrected Hawking
  Radiation from Near-Extremal Kerr-Newman Black Holes}},
  \href{http://arxiv.org/abs/2501.08252}{{\tt arXiv:2501.08252}}.

\bibitem{Lin:2025wof}
G.~Lin, L.~V. Iliesiu, and M.~Usatyuk, {\it {The evaporation of black holes in
  supergravity}},  {\em JHEP} {\bf 08} (2025) 220,
  [\href{http://arxiv.org/abs/2504.21077}{{\tt arXiv:2504.21077}}].

\bibitem{Kapec:2024zdj}
D.~Kapec, Y.~T.~A. Law, and C.~Toldo, {\it {Quasinormal Corrections to
  Near-Extremal Black Hole Thermodynamics}},
  \href{http://arxiv.org/abs/2409.14928}{{\tt arXiv:2409.14928}}.

\bibitem{Kolanowski:2024zrq}
M.~Kolanowski, D.~Marolf, I.~Rakic, M.~Rangamani, and G.~J. Turiaci, {\it
  {Looking at extremal black holes from very far away}},  {\em JHEP} {\bf 04}
  (2025) 020, [\href{http://arxiv.org/abs/2409.16248}{{\tt arXiv:2409.16248}}].

\bibitem{Emparan:2025qqf}
R.~Emparan and S.~Trezzi, {\it {Quantum transparency of near-extremal black
  holes}},  {\em JHEP} {\bf 10} (2025) 023,
  [\href{http://arxiv.org/abs/2507.03398}{{\tt arXiv:2507.03398}}].

\bibitem{Emparan:2025sao}
R.~Emparan, {\it {Quantum cross-section of near-extremal black holes}},  {\em
  JHEP} {\bf 04} (2025) 122, [\href{http://arxiv.org/abs/2501.17470}{{\tt
  arXiv:2501.17470}}].

\bibitem{Biggs:2025nzs}
A.~Biggs, {\it {Following the state of an evaporating charged black hole into
  the quantum gravity regime}},  \href{http://arxiv.org/abs/2503.02051}{{\tt
  arXiv:2503.02051}}.

\bibitem{Betzios:2025sct}
P.~Betzios, O.~Papadoulaki, and Y.~Zhou, {\it {Near-extremal quantum
  cross-section for charged fields and superradiance}},
  \href{http://arxiv.org/abs/2507.13896}{{\tt arXiv:2507.13896}}.

\bibitem{Liu:2024gxr}
X.-L. Liu, J.~Nian, and L.~A. Pando~Zayas, {\it {Quantum Corrections to
  Holographic Strange Metal at Low Temperature}},
  \href{http://arxiv.org/abs/2410.11487}{{\tt arXiv:2410.11487}}.

\bibitem{Heydeman:2024ezi}
M.~Heydeman and C.~Toldo, {\it {The spectrum of near-BPS Kerr-Newman black
  holes and the ABJM mass gap}},  \href{http://arxiv.org/abs/2412.03697}{{\tt
  arXiv:2412.03697}}.

\bibitem{Li:2025vcm}
R.~Li, Z.-X. Man, and J.~Wang, {\it {Decoherence of quantum superpositions in
  near-extremal Reissner-Nordstr{\"o}m black holes with quantum gravity
  corrections}},  {\em JHEP} {\bf 08} (2025) 079,
  [\href{http://arxiv.org/abs/2505.07480}{{\tt arXiv:2505.07480}}].

\bibitem{Castro:2025itb}
A.~Castro, R.~Mancilla, and I.~Papadimitriou, {\it {Near-extremal dynamics away
  from the horizon}},  \href{http://arxiv.org/abs/2507.01126}{{\tt
  arXiv:2507.01126}}.

\bibitem{Maulik:2025phe}
S.~Maulik, A.~Mitra, D.~Mukherjee, and A.~Ray, {\it {Logarithmic corrections to
  near-extremal entropy of charged de Sitter black holes}},
  \href{http://arxiv.org/abs/2503.08617}{{\tt arXiv:2503.08617}}.

\bibitem{Blacker:2025zca}
M.~J. Blacker, A.~Castro, W.~Sybesma, and C.~Toldo, {\it {Quantum corrections
  to the path integral of near extremal de Sitter black holes}},
  \href{http://arxiv.org/abs/2503.14623}{{\tt arXiv:2503.14623}}.

\bibitem{Mariani:2025hee}
F.~Mariani and C.~Toldo, {\it {Gravitational dynamics of near-extreme Kerr
  (Anti-)de Sitter black holes}},  \href{http://arxiv.org/abs/2505.02674}{{\tt
  arXiv:2505.02674}}.

\bibitem{Arnaudo:2025btb}
P.~Arnaudo, G.~Bonelli, and A.~Tanzini, {\it {One loop corrections to the
  thermodynamics of near-extremal Kerr-(A)dS black holes from Heun equation}},
  \href{http://arxiv.org/abs/2506.08959}{{\tt arXiv:2506.08959}}.

\bibitem{Chen:2025lnk}
C.-M. Chen, C.-C. Huang, and S.~P. Kim, {\it {Saturation of Pauli blocking in
  near-extremal charged Nariai black holes}},
  \href{http://arxiv.org/abs/2509.08511}{{\tt arXiv:2509.08511}}.

\bibitem{Montero:2019ekk}
M.~Montero, T.~Van~Riet, and V.~Venken, {\it {Festina Lente: EFT Constraints
  from Charged Black Hole Evaporation in de Sitter}},  {\em JHEP} {\bf 01}
  (2020) 039, [\href{http://arxiv.org/abs/1910.01648}{{\tt arXiv:1910.01648}}].

\bibitem{Higuchi:1986ww}
A.~Higuchi, {\it {Quantization of Scalar and Vector Fields Inside the
  Cosmological Event Horizon and Its Application to Hawking Effect}},  {\em
  Class. Quant. Grav.} {\bf 4} (1987) 721.

\bibitem{Bunch:1978yq}
T.~S. Bunch and P.~C.~W. Davies, {\it {Quantum Field Theory in de Sitter Space:
  Renormalization by Point Splitting}},  {\em Proc. Roy. Soc. Lond. A} {\bf
  360} (1978) 117--134.

\bibitem{Allen:1985ux}
B.~Allen, {\it {Vacuum States in de Sitter Space}},  {\em Phys. Rev. D} {\bf
  32} (1985) 3136.

\bibitem{Chopping:2024oiu}
A.~J. Chopping, C.~Sleight, and M.~Taronna, {\it {Cosmological correlators for
  Bogoliubov initial states}},  {\em JHEP} {\bf 09} (2024) 152,
  [\href{http://arxiv.org/abs/2407.16652}{{\tt arXiv:2407.16652}}].

\bibitem{Higuchi:2018tuk}
A.~Higuchi and K.~Yamamoto, {\it {Vacuum state in de Sitter spacetime with
  static charts}},  {\em Phys. Rev. D} {\bf 98} (2018), no.~6 065014,
  [\href{http://arxiv.org/abs/1808.02147}{{\tt arXiv:1808.02147}}].

\bibitem{Alicki:2023rfv}
R.~Alicki, G.~Barenboim, and A.~Jenkins, {\it {Quantum thermodynamics of de
  Sitter space}},  {\em Phys. Rev. D} {\bf 108} (2023), no.~12 123530,
  [\href{http://arxiv.org/abs/2307.04800}{{\tt arXiv:2307.04800}}].

\bibitem{Page:2012fn}
D.~N. Page and X.~Wu, {\it {Massless Scalar Field Vacuum in de Sitter
  Spacetime}},  {\em JCAP} {\bf 11} (2012) 051,
  [\href{http://arxiv.org/abs/1204.4462}{{\tt arXiv:1204.4462}}].

\bibitem{Bernar:2014lna}
R.~P. Bernar, L.~C.~B. Crispino, and A.~Higuchi, {\it {Infrared-finite graviton
  two-point function in static de Sitter space}},  {\em Phys. Rev. D} {\bf 90}
  (2014), no.~2 024045, [\href{http://arxiv.org/abs/1405.3827}{{\tt
  arXiv:1405.3827}}].

\bibitem{Castro:2022cuo}
A.~Castro, F.~Mariani, and C.~Toldo, {\it {Near-extremal limits of de Sitter
  black holes}},  {\em JHEP} {\bf 07} (2023) 131,
  [\href{http://arxiv.org/abs/2212.14356}{{\tt arXiv:2212.14356}}].

\bibitem{Mertens:2018fds}
T.~G. Mertens, {\it {The Schwarzian theory {\textemdash} origins}},  {\em JHEP}
  {\bf 05} (2018) 036, [\href{http://arxiv.org/abs/1801.09605}{{\tt
  arXiv:1801.09605}}].

\bibitem{Harmark:2007jy}
T.~Harmark, J.~Natario, and R.~Schiappa, {\it {Greybody Factors for
  d-Dimensional Black Holes}},  {\em Adv. Theor. Math. Phys.} {\bf 14} (2010),
  no.~3 727--794, [\href{http://arxiv.org/abs/0708.0017}{{\tt
  arXiv:0708.0017}}].

\bibitem{Spradlin:2001pw}
M.~Spradlin, A.~Strominger, and A.~Volovich, {\it {Les Houches lectures on de
  Sitter space}},  in {\em {Les Houches Summer School: Session 76: Euro Summer
  School on Unity of Fundamental Physics: Gravity, Gauge Theory and Strings}},
  pp.~423--453, 10, 2001.
\newblock \href{http://arxiv.org/abs/hep-th/0110007}{{\tt hep-th/0110007}}.

\bibitem{Olver:2023DLMF}
{NIST Digital Library of Mathematical Functions}, ``{NIST Digital Library of
  Mathematical Functions}.'' \url{https://dlmf.nist.gov/}, 2023.
\newblock Release 1.1.12 of 2023-12-15, National Institute of Standards and
  Technology, Gaithersburg, MD.

\bibitem{Gradshteyn:2014Ryzhik}
I.~S. Gradshteyn and I.~M. Ryzhik, {\em {Table of Integrals, Series, and
  Products}}.
\newblock Academic Press, New York, 8th~ed., 2014.

\end{thebibliography}\endgroup
\end{document}